\documentclass[prc,aps,groupedaddress,superscriptaddress,twocolumn,nofootinbib,showpacs,showkeys,floatfix,a4paper,10pt,noeprint]{
revtex4-1}

\usepackage{graphicx,colordvi,rotating}
\usepackage{multirow,color}
\usepackage{amsmath,amssymb}
\usepackage[mathscr]{euscript}
\usepackage{cancel}
\usepackage{braket}
\usepackage{cases}
\usepackage{color}
\usepackage{mathtools}
\usepackage{natbib}
\usepackage{ulem}
\usepackage{dcolumn}
\usepackage{bm}
\usepackage{natbib}
\usepackage{enumerate}
\usepackage[dvipsnames]{xcolor}
\usepackage{wrapfig}
\usepackage[colorlinks,citecolor=blue]{hyperref}

\usepackage[utf8]{inputenc}
\usepackage{epstopdf}

\usepackage{t1enc}
\usepackage[utf8]{inputenc}

\begin{document}


\title{Microscopic pairing in fission dynamics}

\author{A. Zdeb}
\email{azdeb@kft.umcs.lublin.pl}
\affiliation{Institute of Physics, Maria-Curie-Skłodowska Univeristy, Lublin, Poland}

\author{A. Baran}
\email{barandrzej@gmail.com}
\affiliation{Institute of Physics, Maria-Curie-Skłodowska Univeristy, Lublin, Poland}

\author{S.A. Giuliani}
\email{samuel.giuliani@uam.es}
\affiliation{Departamento de Física Teórica and CIAFF, Universidad Autónoma de Madrid, Madrid, Spain}

\author{L.M. Robledo}
\email{luis.robledo@uam.es}
\affiliation{Departamento de Física Teórica and CIAFF, Universidad Autónoma de Madrid, Madrid, Spain}
\affiliation{Center for Computational Simulation, Universidad Politécnica de Madrid, Campus de Montegancedo, Bohadilla 
del Monte, E-28660 Madrid, Spain}

\author{M. Warda}
\email{warda@kft.umcs.lublin.pl}
\affiliation{Institute of Physics, Maria-Curie-Skłodowska Univeristy, Lublin, Poland}

\date{\today}

\begin{abstract}
Nuclear fission can be modelled as a quantum tunneling process driven by the interplay between the nuclear binding energy and the collective inertia. Within the Wentzel-Kramers-Brillouin formalism, spontaneous fission half-lives can be obtained by minimizing the action integral in the multidimensional space of collective degrees of freedom. Hence, including the relevant collective variables is crucial for properly describing spontaneous fission probabilities. 
Pairing correlations play an essential role in this evaluation since the collective 
inertia decreases as the inverse of the square of the pairing gap, and, therefore, they should be considered as a relevant degree of freedom on the same footing as deformation parameters.
In this work, we show that the spontaneous fission half-lives in fermium isotopes can be reproduced in a microscopic theory 
by considering the least-action fission path in a two-dimensional space with constraints on the quadrupole moment and 
pairing correlations. We consider two microscopic quantities as degrees of freedom associated with pairing: the pairing gap parameter $\Delta$, and the particle number fluctuations $\langle \Delta 
N^2 \rangle$. Least-action paths, computed using the Dijkstra algorithm, are compared with minimum-energy paths, highlighting the importance of pairing correlations as a dynamical degree of freedom.
\end{abstract}


\maketitle


\section{Introduction}

Spontaneous fission is the dominant decay mode in many heavy and super-heavy nuclei. A relevant observable describing 
this process is half-life, which can be directly linked to the probability of the nucleus to tunnel through the fission barrier that can be estimated by using, for instance, the semi-classical Wentzel-
Kramers-Brillouin (WKB) method~\cite{messiah1961quantum}. Currently, the reproduction of experimental spontaneous fission lifetimes is still a challenge in theoretical nuclear physics, as the description of fission 
requires both a deep understanding of the properties of the fission barrier created by nuclear forces and the process of 
tunneling through it. The traditional evaluation of spontaneous fission half-lives is based on determining the potential energy surface (PES) in the multidimensional deformation space built from constraints 
or deformation parameters. The fission barrier is computed along the least-energy path that minimizes energy on the way 
from the ground state to the scission point.  Such an approach is called {\it 
static}, as a single fission path is determined by simple minimization of the energy for each elongation parameter 
independently.

On its way to fission, however, the nucleus will dominantly follow the path which maximizes the tunneling probability, minimizing the action integral: the so-called least-action path. Because action depends both on energy and collective 
inertia, the path should be determined in a {\it dynamic} way, searching for an optimal combination of the two 
quantities from ground state to scission step by step. Contrary to the static path, where the relevant degrees of freedom are those associated with a reduction of the fission barrier, in dynamical calculations, it is crucial to include collective degrees of freedom that can reduce collective inertia.

Pairing correlations between nucleons play an important role in fission, as they govern the level density around the 
Fermi level, which impacts the height of the fission barrier and, to a large extent, the collective inertias. As it was 
shown over fifty years ago, collective inertias decrease as the inverse of the square of the pairing gap 
\cite{Brack1972,LEDERGERBER19731,babinet,Jensen_1983}. This reduction in collective inertias, resulting from 
enhanced pairing correlations, may lead to an alternative trajectory: one with a larger fission 
barrier, but an increased tunneling probability. Thus, a precise description of the dynamic fission path should include not only
the potential energy surface in the deformation space, but also collective degrees of freedom associated with pairing correlations, such as the pairing gap $\Delta$ or the particle number fluctuation $\langle \Delta N^2 \rangle$.
Such calculations have been performed in macroscopic-microscopic 
models~\cite{BARAN1978,STASZCZAK1985227,STASZCZAK1989589,LOJEWSKI1999} and later in the microscopic self-consistent 
calculations~\cite{Giuliani2014, PhysRevC.90.061304, PhysRevC.93.044315, PhysRevC.107.044307} (for a recent
literature review on the impact of pairing correlations on fission, see Ref.~\cite{zdeb2025}).

In this work, we explore the impact of the dynamically evaluated pairing correlations on the calculation of spontaneous 
fission half-lives. We determine the quantities entering the action integral using the self-consistent Hartree-Fock-Bogoliubov (HFB) theory over a two-dimensional collective space defined by the quadrupole moment 
and the pairing degree of freedom. The latter is set by either the pairing gap parameter $\Delta$ or the average particle number 
fluctuations $\langle \Delta N^2 \rangle$. 
We study the properties of the dynamic fission path in $^{262}$Rf, and the evolution of fission barriers and half-lives along the fermium isotopic chain, whose characteristic inverted parabola
of experimental half-lives represents a challenge for any theoretical model. The half-lives obtained from dynamical calculations show significantly better agreement with experimental data than those based on static least-energy paths, demonstrating the importance of including
dynamic pairing correlations in spontaneous fission studies.

\section{Theoretical framework \label{THEORY}}
Spontaneous fission properties are determined by the nucleus tunneling process through the fission barrier on its way to scission. As mentioned in the introduction, the tunneling probability can be estimated by
using the WKB semi-classical approach, depending on the
exponential of the classical action defined with respect to the collective
variables. The shortest-lived 
fission channel (i.e., main fission decay channel) 
is thus the one minimizing the action, which  
depends on the inertia of collective motion and the energy difference 
between the ground state and the energy at the barrier. If the inertia 
weakly depends on the collective coordinates characterizing fission, the path of least action is equivalent to the path of least-energy. This is a widely used assumption in many fission models estimating spontaneous fission half-lives, as the least-energy path can be directly obtained from a variational principle.
However, inertia strongly depends on the amount of pairing
correlations present in the process, and it has to be included as a collective
variable playing a leading role in fission dynamics. In this work, we characterize the fission properties using a
least-action scheme with both quadrupole deformation and pairing
as fission coordinates.

To compute the quantities entering the action integral, we use the microscopic, self-consistent 
HFB model with the finite-range Gogny-type nuclear interaction D1S~\cite{D1S1,D1S2,D1S3,D1S4,D1S5}. The HFB model allows for a consistent calculation of both the PES and 
the collective inertia required to compute the action. The PES ($V$) is defined 
as the set of HFB energies of a constrained calculation supplemented with the standard rotational energy correction computed in the strong deformation limit~\cite{Egido2004}, and the zero-point energy correction stemming from vibrational fluctuations in the quadrupole degree of freedom. Coulomb exchange is computed within the Slater approximation with the neglect of Coulomb antipairing~\cite{PhysRevC.99.064301}. The two-body kinetic energy correction is taken into account in the variational process. The collective inertia ($B$) is computed in the so-called Adiabatic
Time Dependent HFB (ATDHFB) framework with the ``perturbative cranking'' approximation~\cite{Yuldashbaeva1999,PhysRevC.84.054321,Giuliani2018b}.

The method used to solve the HFB equation is the gradient
method supplemented with an approximate curvature correction~\cite{Robledo2011d}. Apart from the reduced number of iterations, one of the advantages of this method is its ability to deal with
all kinds of constraints, including two-body operator constraints. 
Axial symmetry is preserved in the HFB calculation, as previous studies have shown that the 
least-action principle tends to drive the system away from triaxial configurations~\cite{PhysRevC.90.061304}. On the other hand, reflection symmetry is allowed to be broken. The quasiparticle operators of the Bogoliubov transformation are expanded in a one-center, axially symmetric, deformed harmonic oscillator basis with 16 deformed shells ($n_z=0,\ldots,22$) 
and the two oscillator lengths parameters of the basis states are optimized for each of 
the considered HFB wave functions. 

As mentioned before, the calculation of the spontaneous fission half-life uses the WKB approximation
where the probability of tunneling through the one-dimensional fission barrier is the exponential
of the action integral computed in the classically forbidden region under the barrier. The action
$S[L(s)]$ computed along the fission path $L(s)$ is thus given by 
\begin{equation}
S=\frac{1}{\hbar}\int_\textup{in}^\textup{out}\sqrt{2B_{\rm eff}(s)(V(s)-E_{gs})}ds \,,
\label{s}
\end{equation}
being $in$ and $out$ the classical turning points defined by the condition $V(s)-E_{gs}=0$. 
The ``ground state energy'' $E_{gs}$ 
is the sum of $V(s)$ at the ground state plus the collective energy $E_0$ stemming from the quantization of the quadrupole collective motion. The $E_0$ can
be estimated by considering the curvature of $V(s)$ around the ground state minimum and the
collective inertia in that configuration~\cite{Schindzielorz2009}. However, its value is usually of the order of 0.5 MeV, and we
have taken this reference value for all the calculations performed. 
The definition of the element of length $ds$ depends on the number of collective variables used in the calculations. For a single collective degree of freedom, $s$ corresponds to the collective coordinate (usually the 
axial quadrupole moment $Q_{20}$, or the $\beta_2$ deformation parameter). In higher-dimensional
calculations, $ds$ is the element of length parameterizing the path $L$ in the multidimensional space of collective
variables. In our case, the path across the two-dimensional space is chosen to minimize
the action integral of Eq.~(\ref{s}).
Once the least-action path $L(s)$ is defined and the action $S$ computed, the spontaneous fission half-life (in seconds) is calculated using the standard formula~\cite{Nilsson1969a}:
\begin{equation}
t_{sf}=2.87\times 10^{-21}(1 + \exp(2S)) \,. 
\label{t}
\end{equation}
In the multidimensional case, the effective inertia 
$B_{\rm eff}(s)$ is given by:
\begin{equation}
B_{\rm eff}(s)=\sum_{i,j}B_{ij}\frac{dq_i}{ds}\frac{dq_j}{ds}.
\label{b_eff}
\end{equation}
where the $B_{ij}$ are the inertias for the collective variables $q_i$ and $q_j$ used in the dynamical calculation. 

In this work, we consider a two-dimensional potential energy surface 
defined by the axial quadrupole moment $Q_{20}$ and a degree of freedom associated with pairing 
correlations. The pairing coordinate is obtained by constraining either 
the one-body pairing gap operator $\hat{P}^{+}=\sum_k \ c^+_k c^+_{\bar{k}} 
$ or the two-body particle number fluctuation operator  $ \Delta 
\hat{N}^2=\hat{N}^{2}-\langle \hat{N} \rangle^{2}$. In both cases we constraint 
on the total value obtained by adding the contribution from protons and
neutrons. Treating protons and neutrons separately is still rather challenging
as it implies using three collective coordinates at the same time. 
In spite of being a very interesting issue we have to postpone its study to a
future publication. In systems without pairing, the particle number fluctuation is zero, 
while in nuclei exhibiting strong pairing correlations, it takes on 
noticeably larger values, typically on the order of a few units.

\begin{figure}[!htb]
\includegraphics[width=\columnwidth, angle=0]{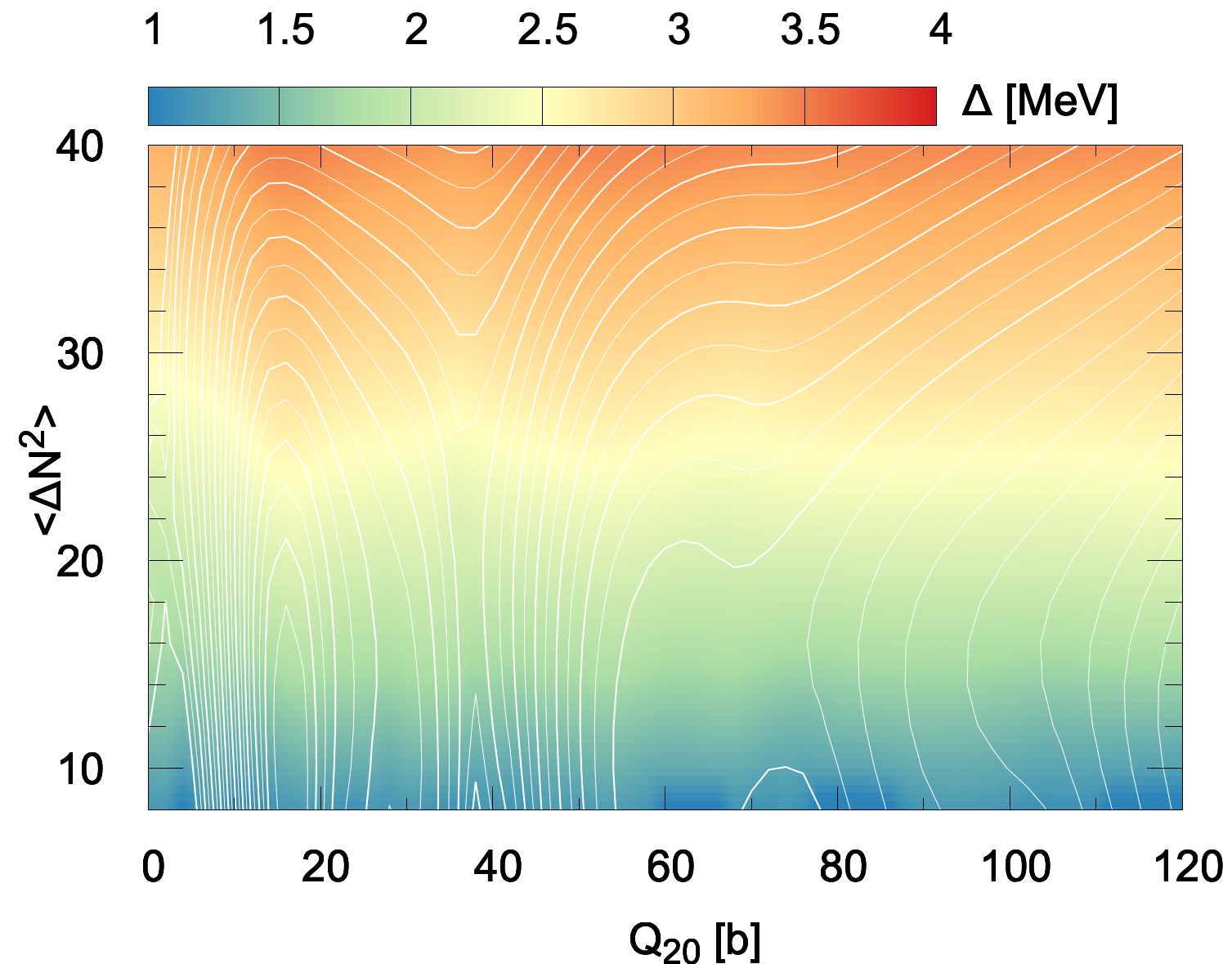}
\caption{Pairing gap ${\Delta}$ as a function of $Q_{20}$ and $\langle \Delta N^2 \rangle$ for $^{262}$Rf obtained with the Gogny D1S 
parametrization. The white isolines correspond to the HFB energy.}
\label{deln}
\end{figure}

In a previous study~\cite{zdeb2025}, we showed that the usage of 
$\Delta=G\langle \hat{P}^{+} \rangle $ ($G=10/A$  is the standard pairing strength
with $A$ the mass number) as a collective degree of freedom can be problematic due to 
the indeterminacy of the sign of the occupation amplitudes $v$ and $u$ 
in the canonical basis. 
On the other hand, the standard expressions for the collective 
inertias are only defined for
one-body operators, preventing their use with the two-body operator  $ \Delta \hat{N}^2 $. The solution to this
problem, as suggested in Ref.~\cite{zdeb2025}, is to perform the constraint calculations with the constraint on 
$\langle \Delta N^2 \rangle$ and for each of the computed HFB wave functions, evaluate the value of $\Delta$ and
the corresponding inertias with the $\hat{P}^{+}$ operator. It was demonstrated that $\Delta$ and 
$\langle \Delta N^2 \rangle$ are, to a very good extent, proportional to each other. 
It was also proven that the inertia computed as if  $\Delta \hat{N}^2$ were a one-body operator is also roughly proportional 
to the inertia computed with $\hat{P}^{+}$. In Fig.~\ref{deln}, a two-dimensional color plot showing the values
of $\Delta$ as a function of $\langle \Delta N^2 \rangle$ for all possible $Q_{20}$ values relevant for 
fission is given for a typical nucleus in the region. The plot shows the proportionality between 
$\langle \Delta N^2 \rangle$ and $\Delta$ and how the proportionality constant is nearly independent of $Q_{20}$
except for configurations near sphericity. In the next Section, we will present results obtained
with either $\langle \Delta N^2 \rangle$ or $\Delta$ as pairing-like coordinate to validate the findings
of~\cite{zdeb2025} regarding the equivalence of both degrees of freedom for least-action calculations.

To determine the least-action path in the two-dimensional PES we employ the Dijkstra algorithm~\cite{Dijkstra,clr,sw}.
Dijkstra's algorithm is a popular method used in computer science and graph theory to find the shortest path between two nodes in 
a graph. The graph can represent various types of networks, such as road systems, communication networks, or even 
social networks. The advantage of Dijkstra's algorithm lies in its simplicity and efficiency for solving single-source 
shortest path problems, where one wants to find the shortest path from a given start node to the 
pre-defined final nodes in the PES. The algorithm makes the locally 
optimal choice at each step (choosing the closest unvisited node) to eventually find the global optimum (the shortest 
path). The Dijkstra algorithm has already been applied in fission studies searching the least-action path 
in a two-dimensional space of "geometric" nuclear collective degrees of freedom - quadrupole and 
octupole moment~\cite{PhysRevC.98.024623}. The ground state is set as a start node, and the final nodes are exit points from the two-dimensional fission barrier.
The nodes of the graph are created from the two-dimensional regular, rectangular mesh obtained in the
constrained calculation considering the collective variables $(Q_{20},\langle \Delta N^2 \rangle)$ or $(Q_{20}, \Delta)$. 
We consider that a given node is connected only to the neighboring 
nodes, and the action is treated as a distance between nodes. In this 
work, we have adapted Burkardt's program~\cite{jb} to implement 
Dijkstra's algorithm.

\section{RESULTS}
%

\begin{figure*}[!htb]
\includegraphics[scale=0.130,angle=0]{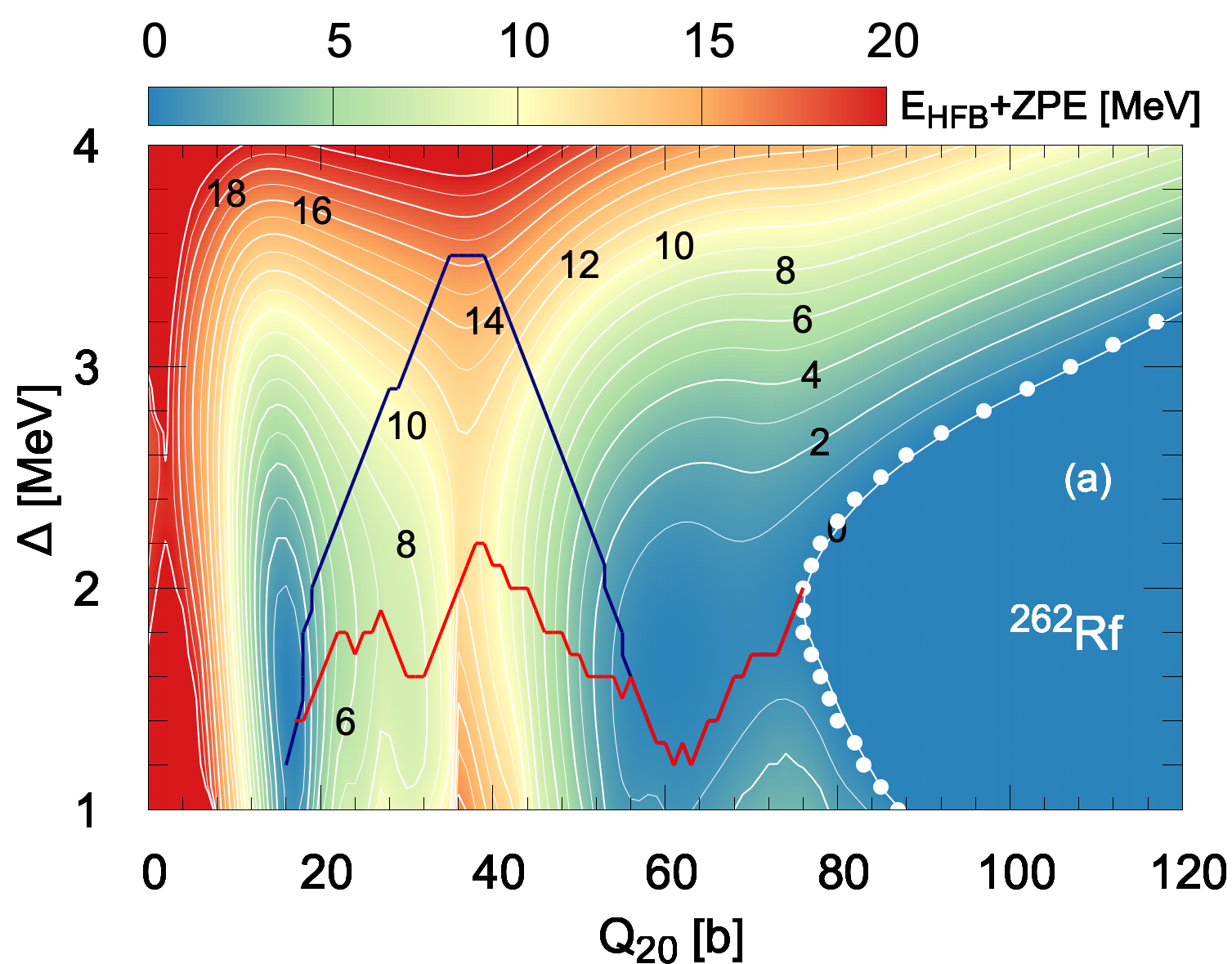}
\includegraphics[scale=0.130,angle=0]{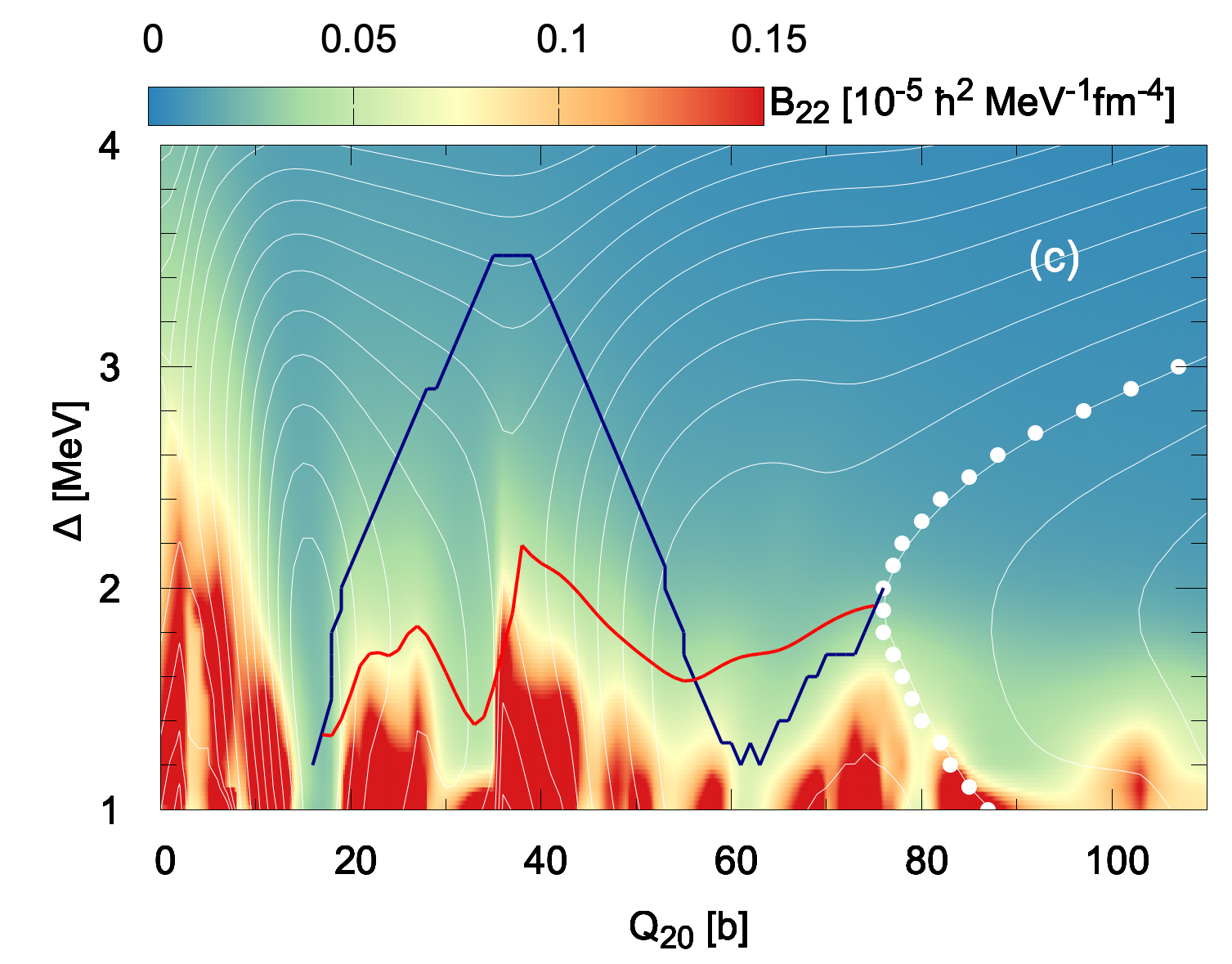}\\
\includegraphics[scale=0.130,angle=0]{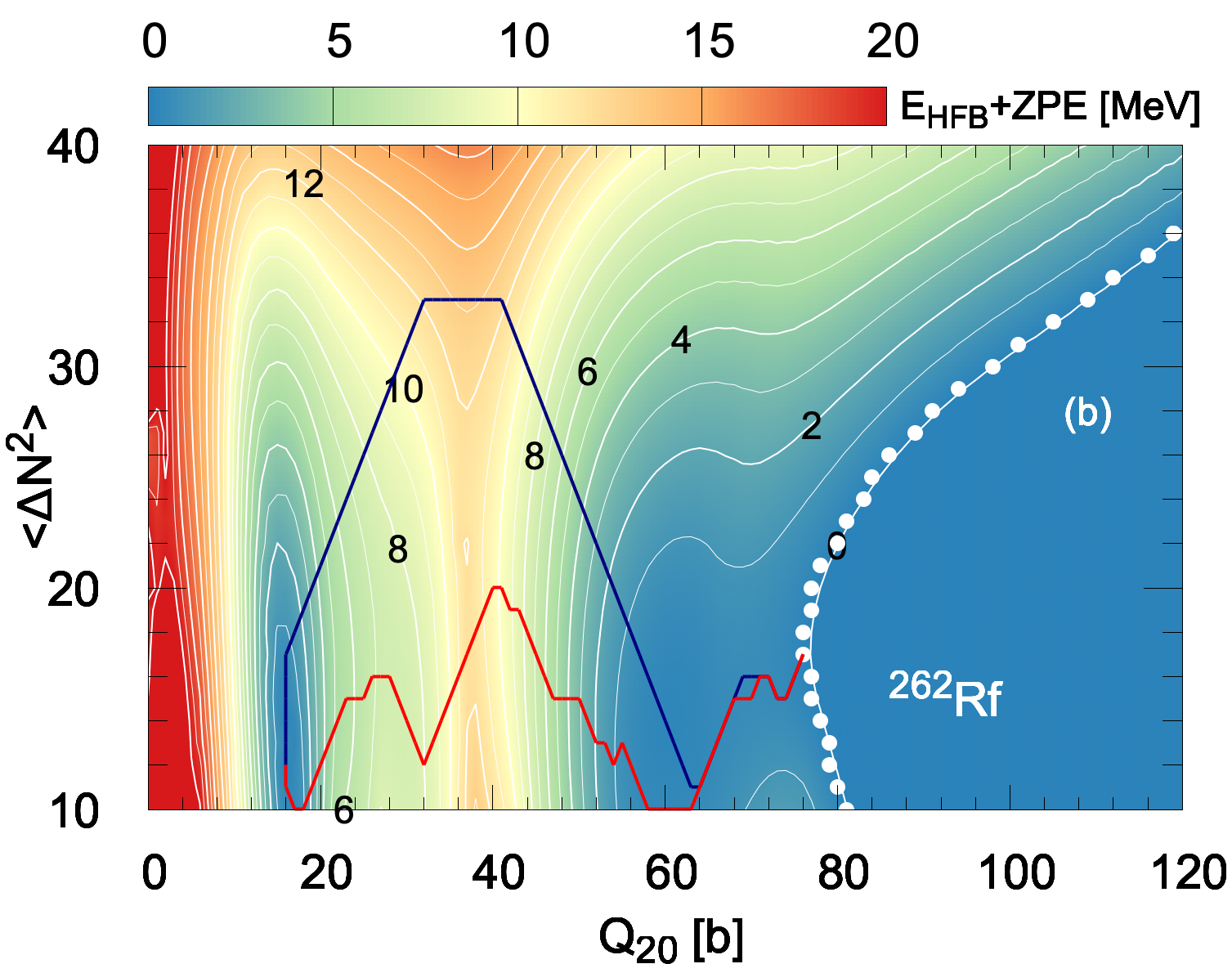}
\includegraphics[scale=0.130,angle=0]{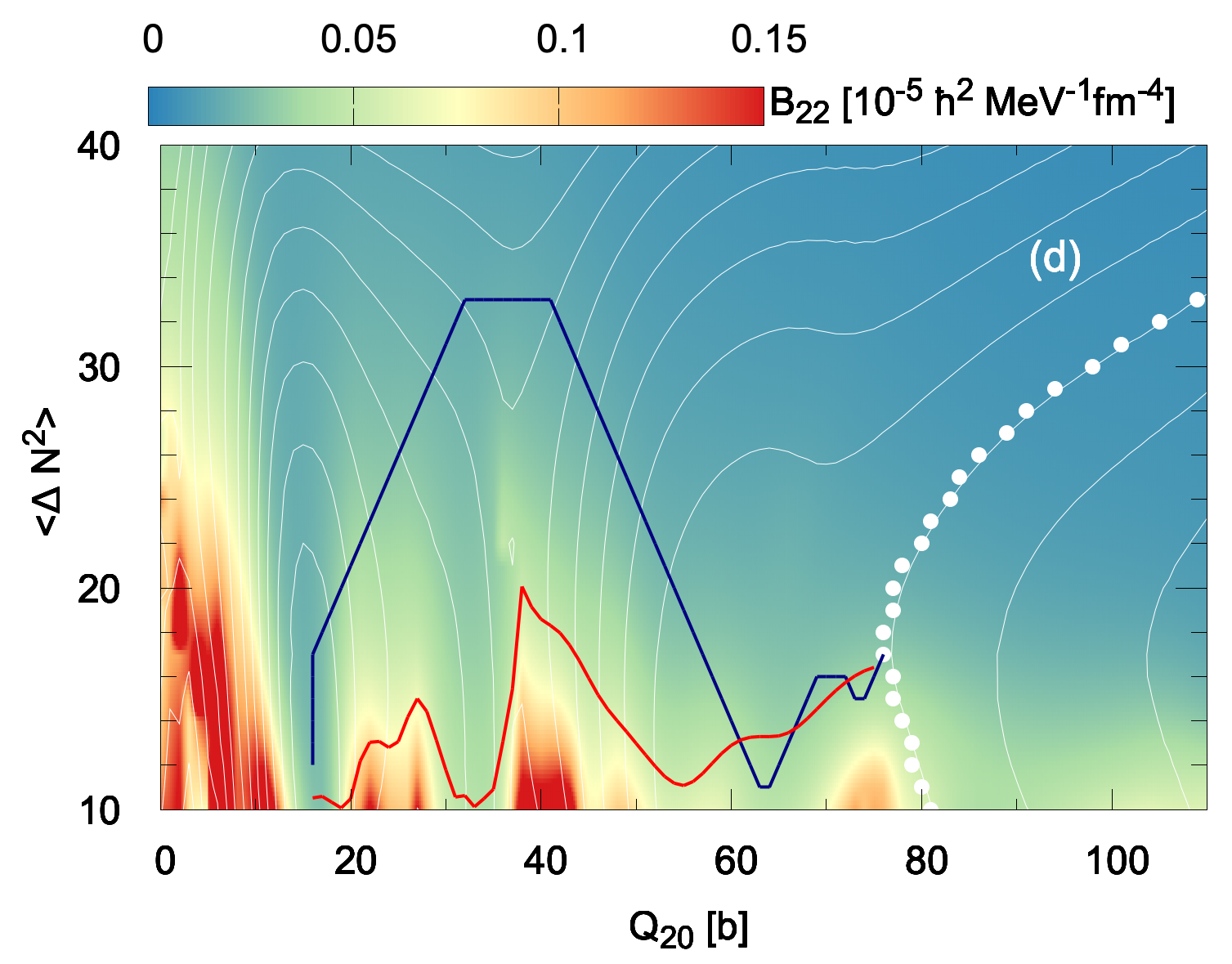}
\caption{PES in MeV (left column) and $B_{22}$ inertia tensor component (right column) in the ($Q_{20}$, $\Delta$) plane (top row) and in the 
($Q_{20}$, $\langle \Delta N^2 \rangle$) plane (bottom row) for $^{262}$Rf, obtained with the Gogny D1S parametrization. The red line corresponds to the 
least-energy path, while the blue one corresponds to the least-action path. The white dots show all the considered exit points from 
the fission barrier.}
\label{dynm}
\end{figure*}

\begin{figure}[!htb]
\includegraphics[width=\columnwidth, angle=0]{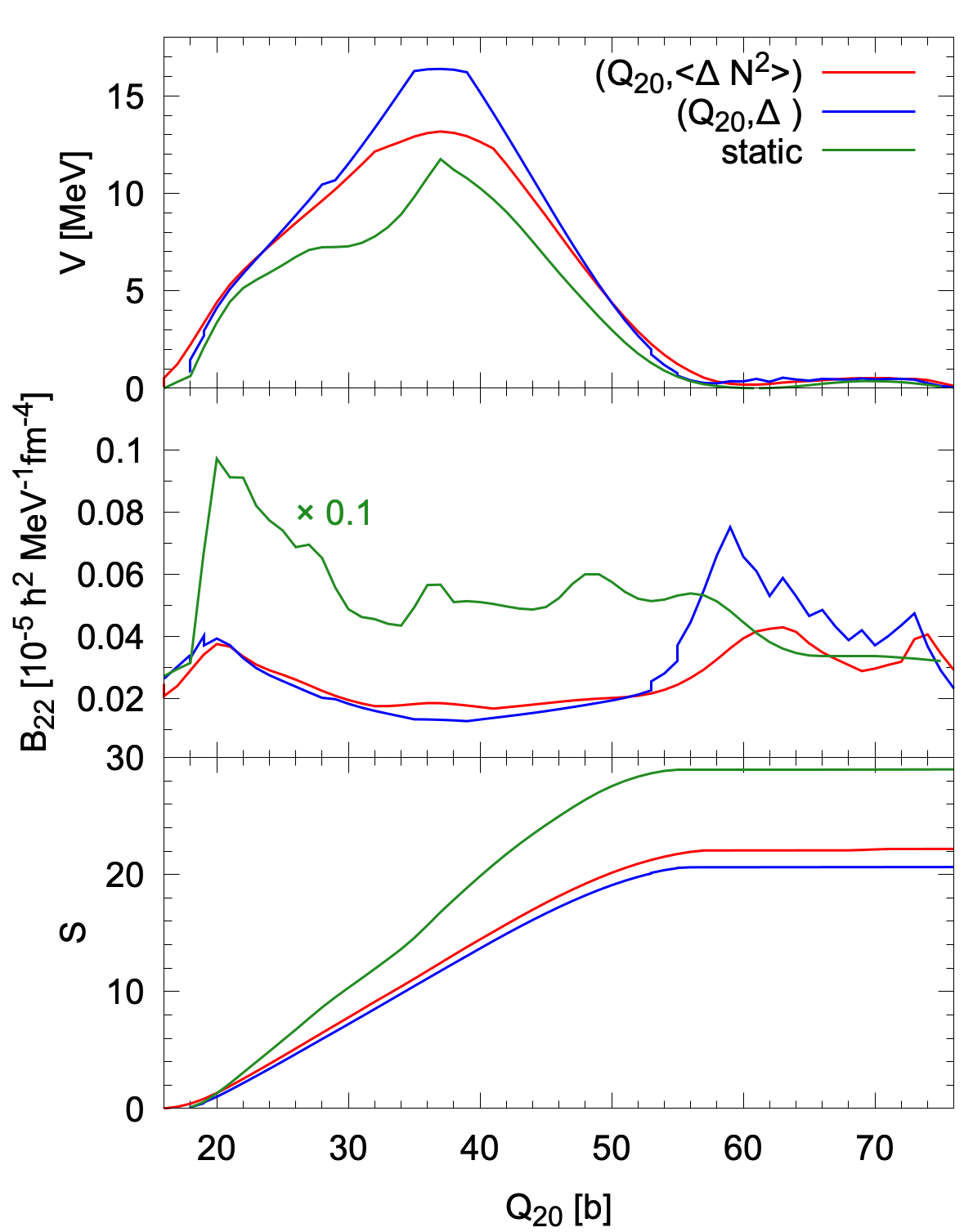}

\caption{The potential energy (top), the $B_{22}$ component of the collective inertia (middle), and the action integral (bottom) as a function of $Q_{20}$ along the least-action fission paths in a ($Q_{20}$, $\langle \Delta N^2 \rangle$) plane (red line) and in a ($Q_{20}$, $\Delta$) (blue line), and the least-energy (static) path (green line) in $^{262}$Rf.
The $B_{22}$ component of the collective inertia for the least-energy (static) path is reduced by a factor of 10.}
\label{1d}
\end{figure}

\begin{figure*}[!htb]
\includegraphics[scale=0.130,angle=0]{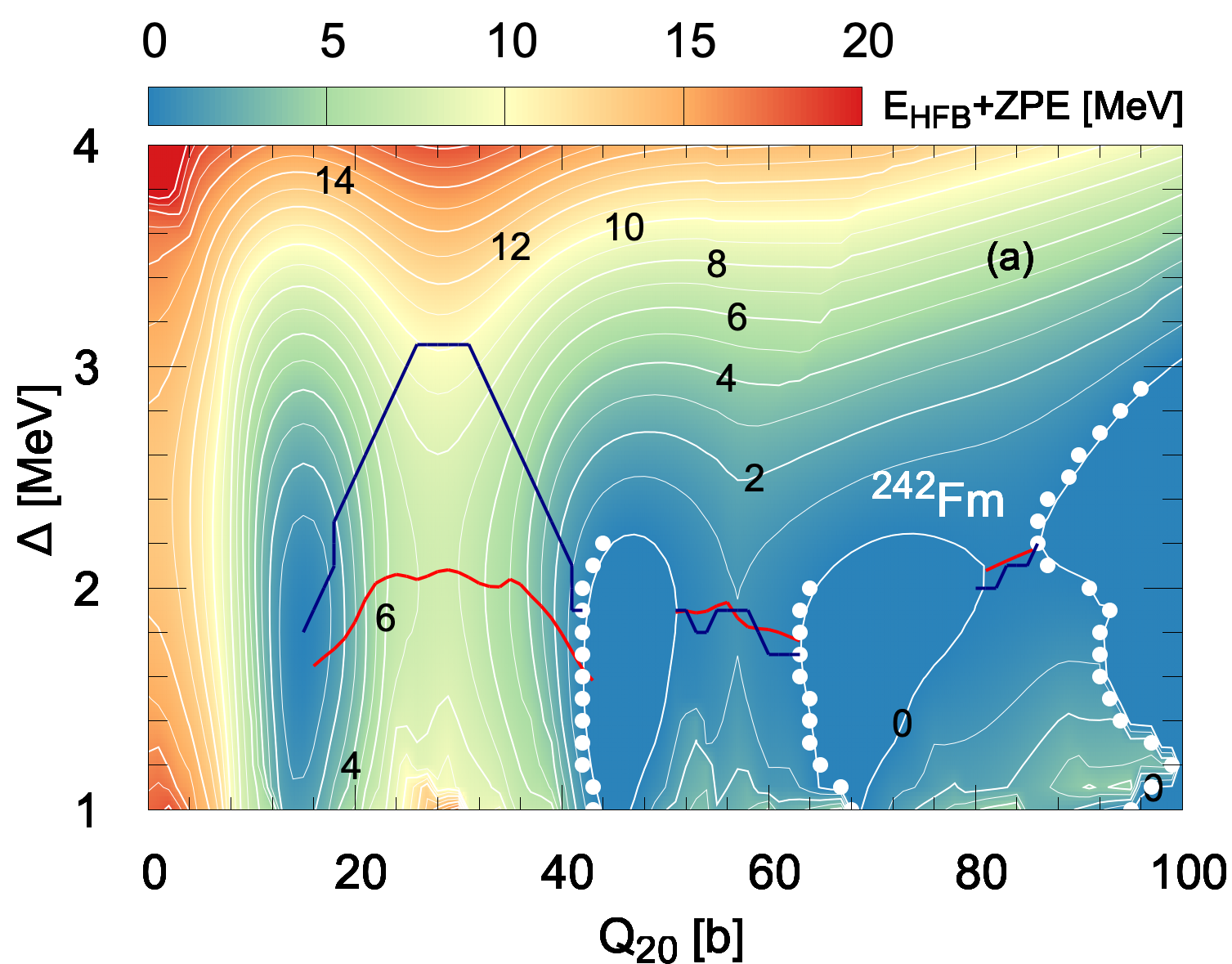}
\hskip-0.1cm
\includegraphics[scale=0.130,angle=0]{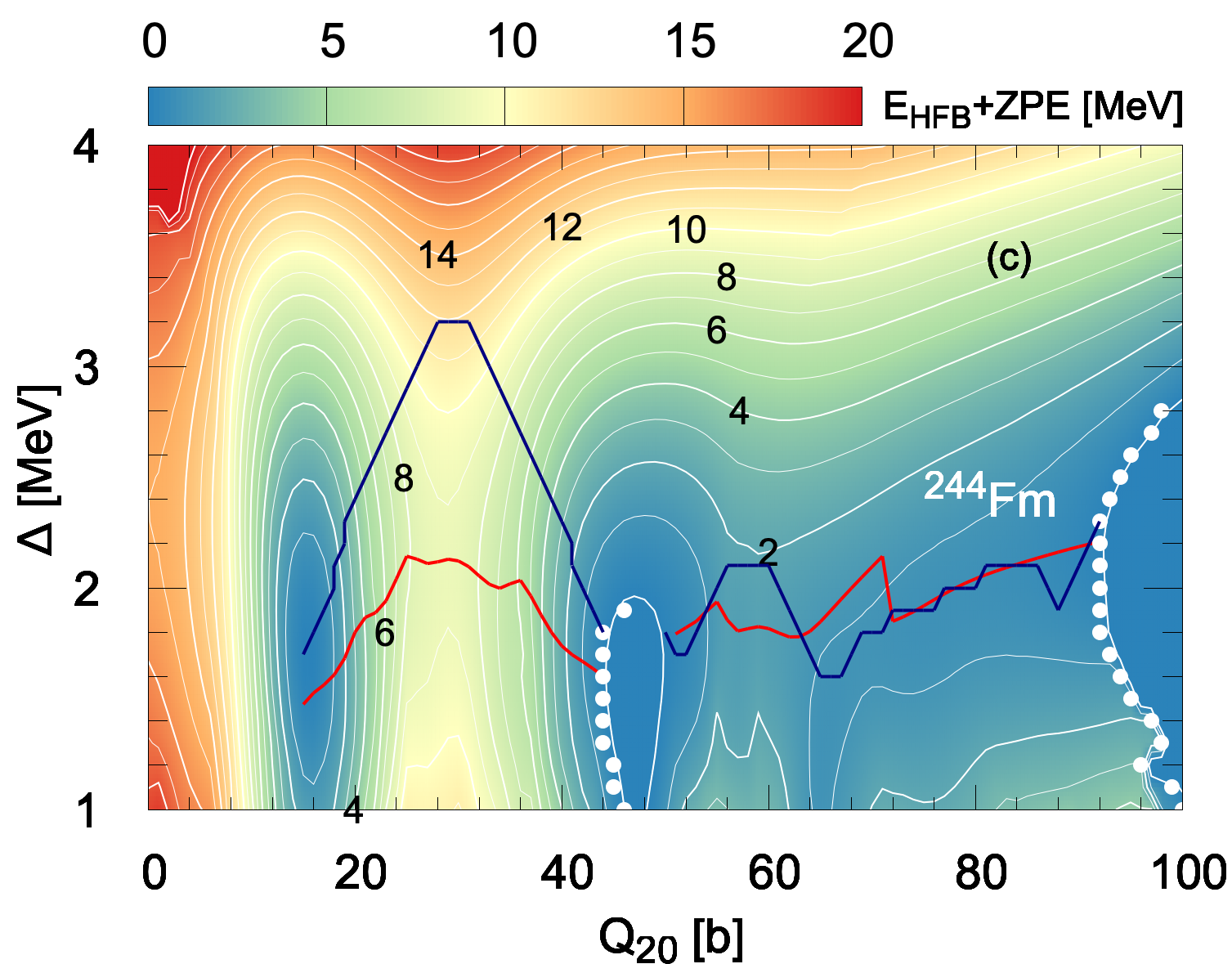}\\
\vskip-0.67cm
\includegraphics[scale=0.130,angle=0]{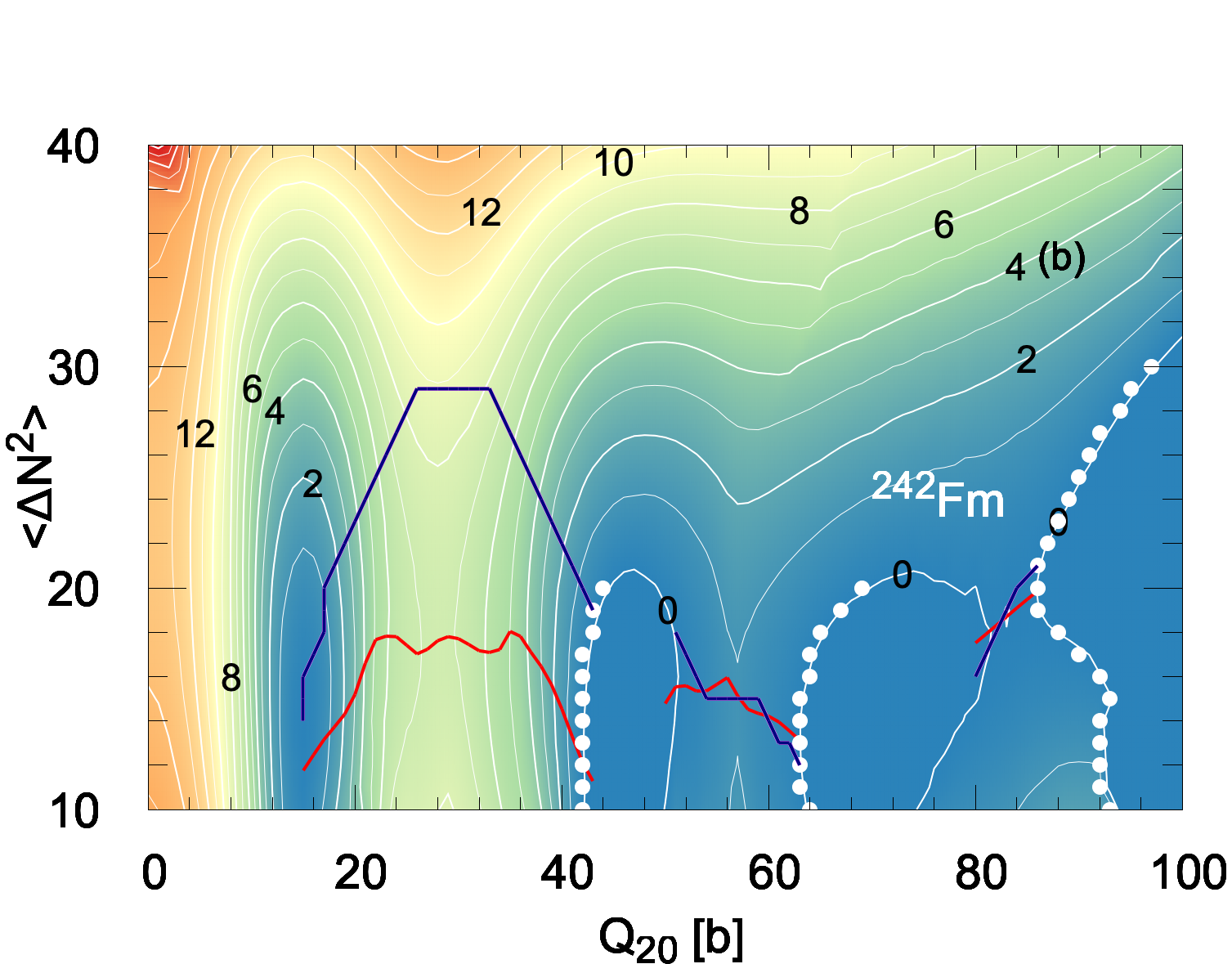}
\hskip-0.1cm
\includegraphics[scale=0.130,angle=0]{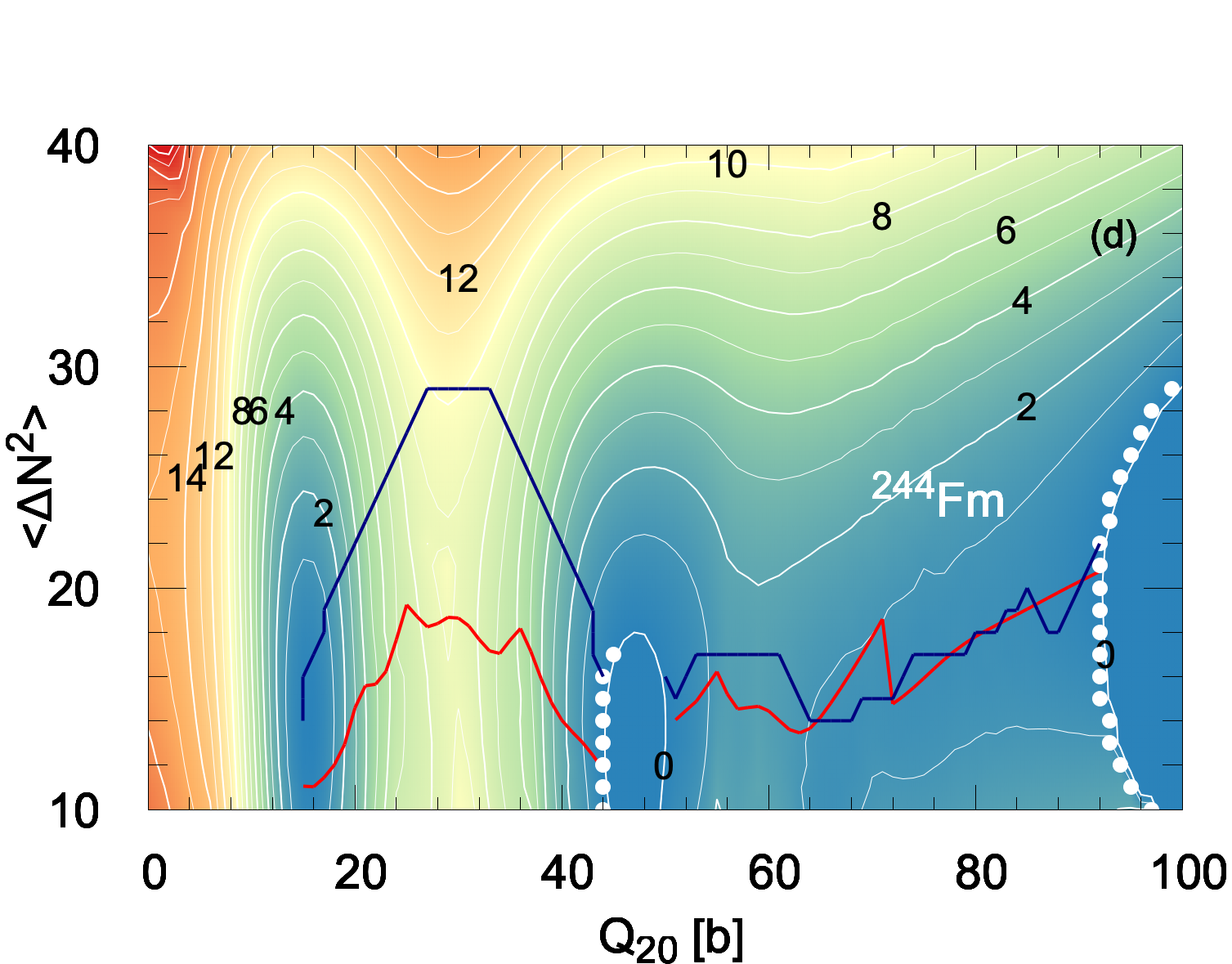}\\
\vskip0.cm
\caption{PES (in MeV) of $^{242}$Fm (left column) and $^{244}$Fm (right column) in the ($Q_{20}$, $\Delta$) plane  (top row) 
($Q_{20}$, $\langle \Delta N^2 \rangle$) (bottom row). The red line corresponds to the 
least-energy path, while the blue line to the least-action path. The white dots show the considered exit points from 
the fission barrier.}
\label{dynFm1}
\end{figure*}

\begin{figure*}[!htb]
\includegraphics[scale=0.130,angle=0]{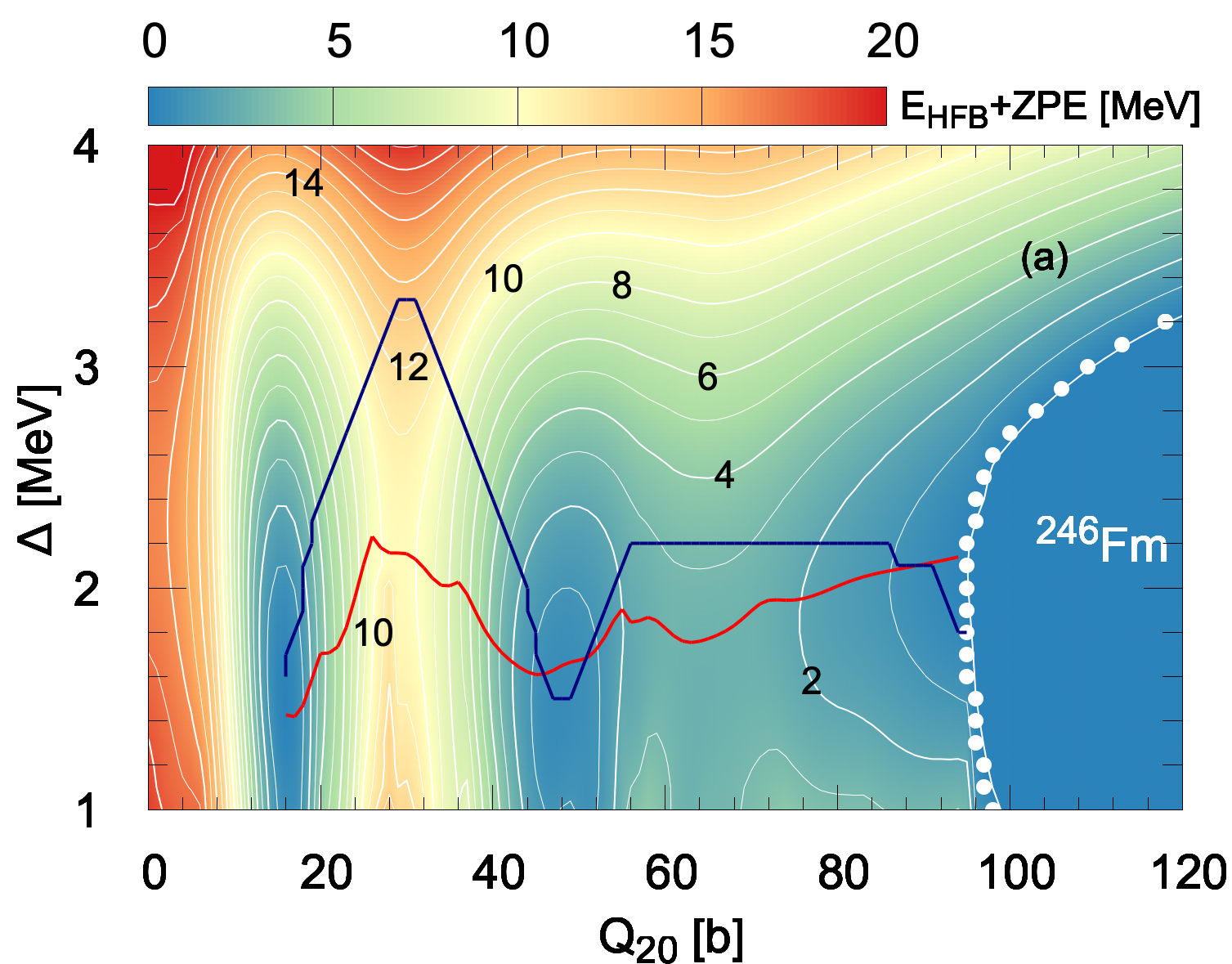}
\hskip-0.1cm
\includegraphics[scale=0.130,angle=0]{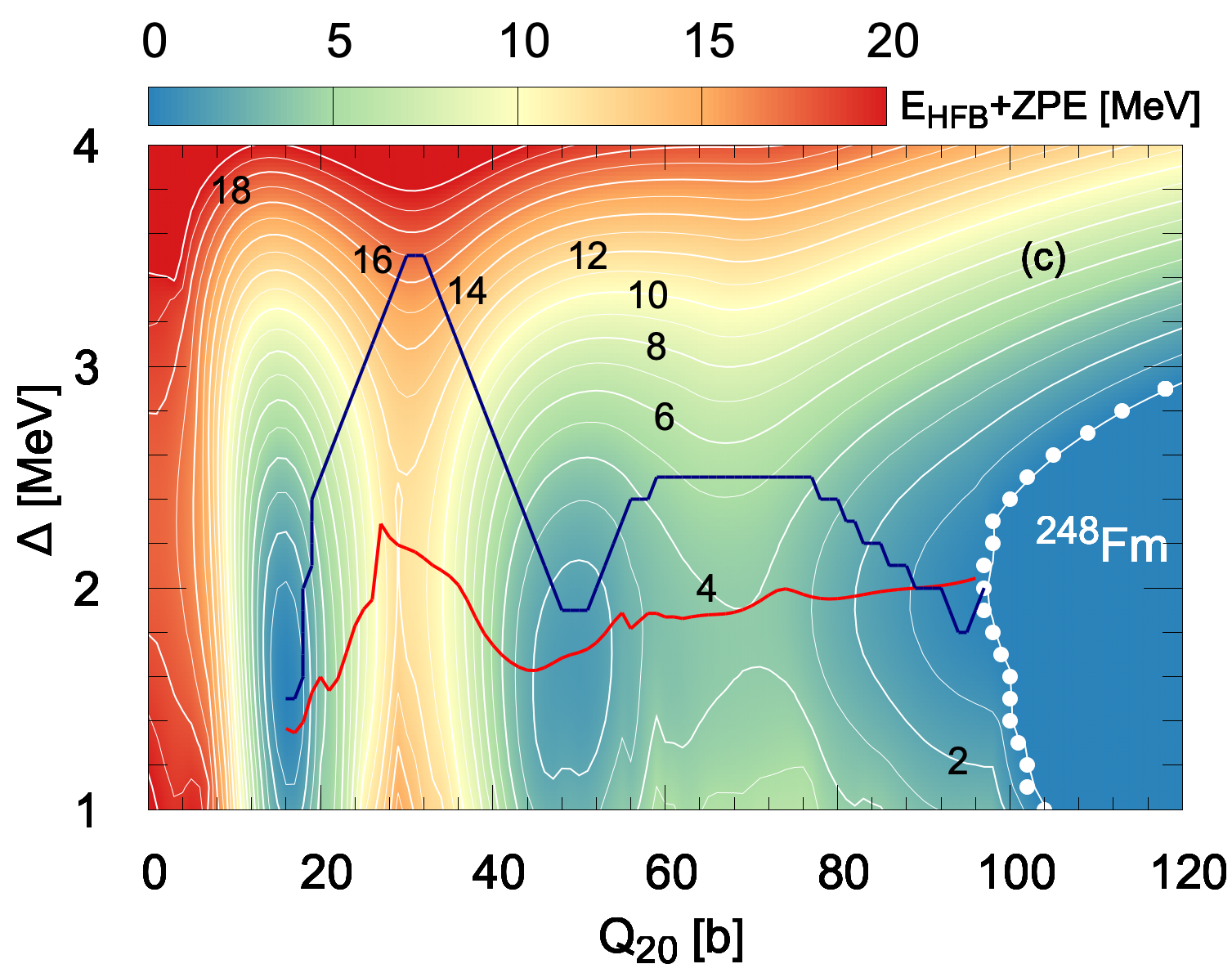}\\
\vskip-0.67cm
\includegraphics[scale=0.130,angle=0]{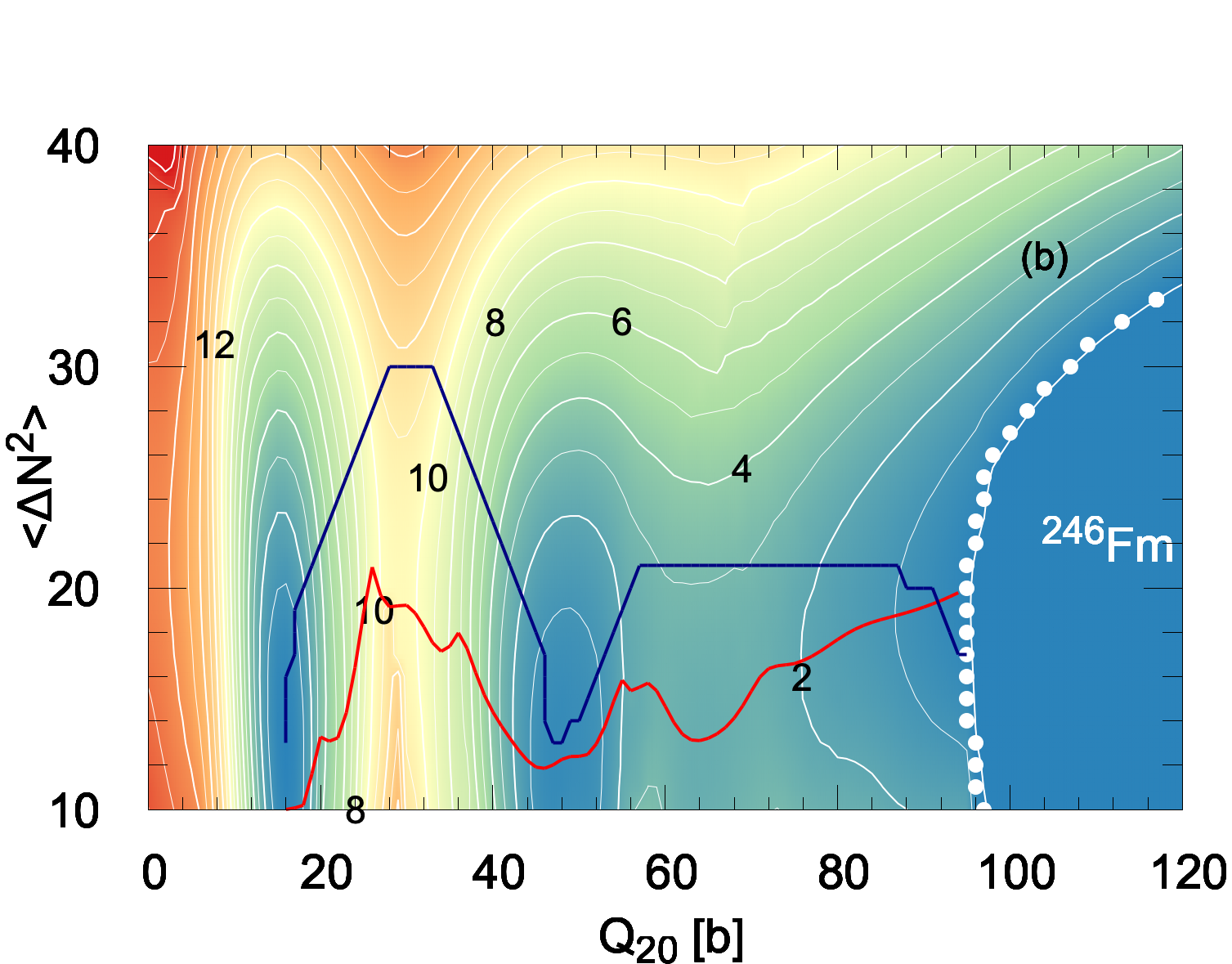}
\hskip-0.1cm
\includegraphics[scale=0.130,angle=0]{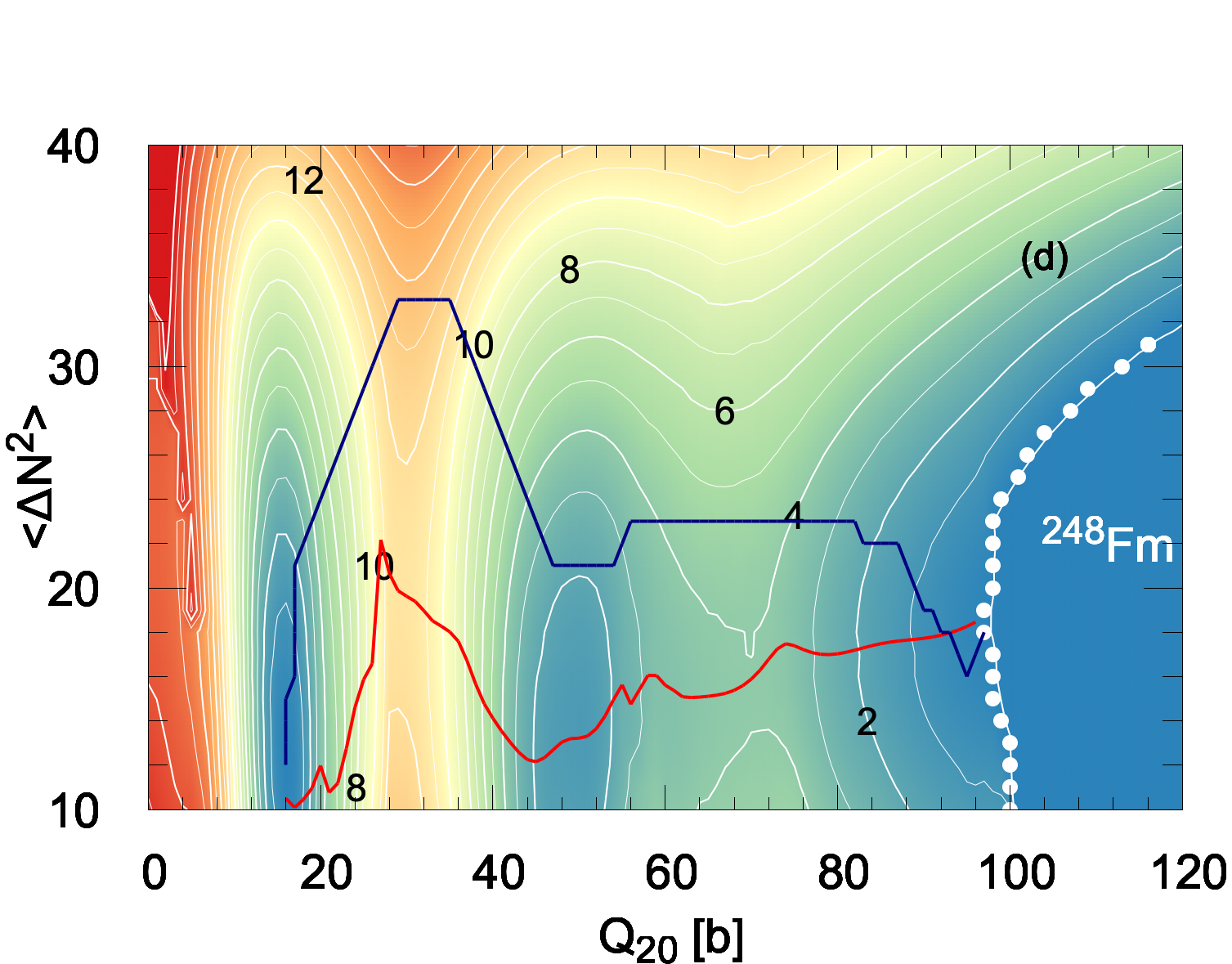}\\
\vskip0.1cm
\caption{Same as Fig.~\ref{dynFm1}, but for $^{246}$Fm and $^{248}$Fm.}
\label{dynFm2}
\end{figure*}
\begin{figure*}[!htb]
\includegraphics[scale=0.130,angle=0]{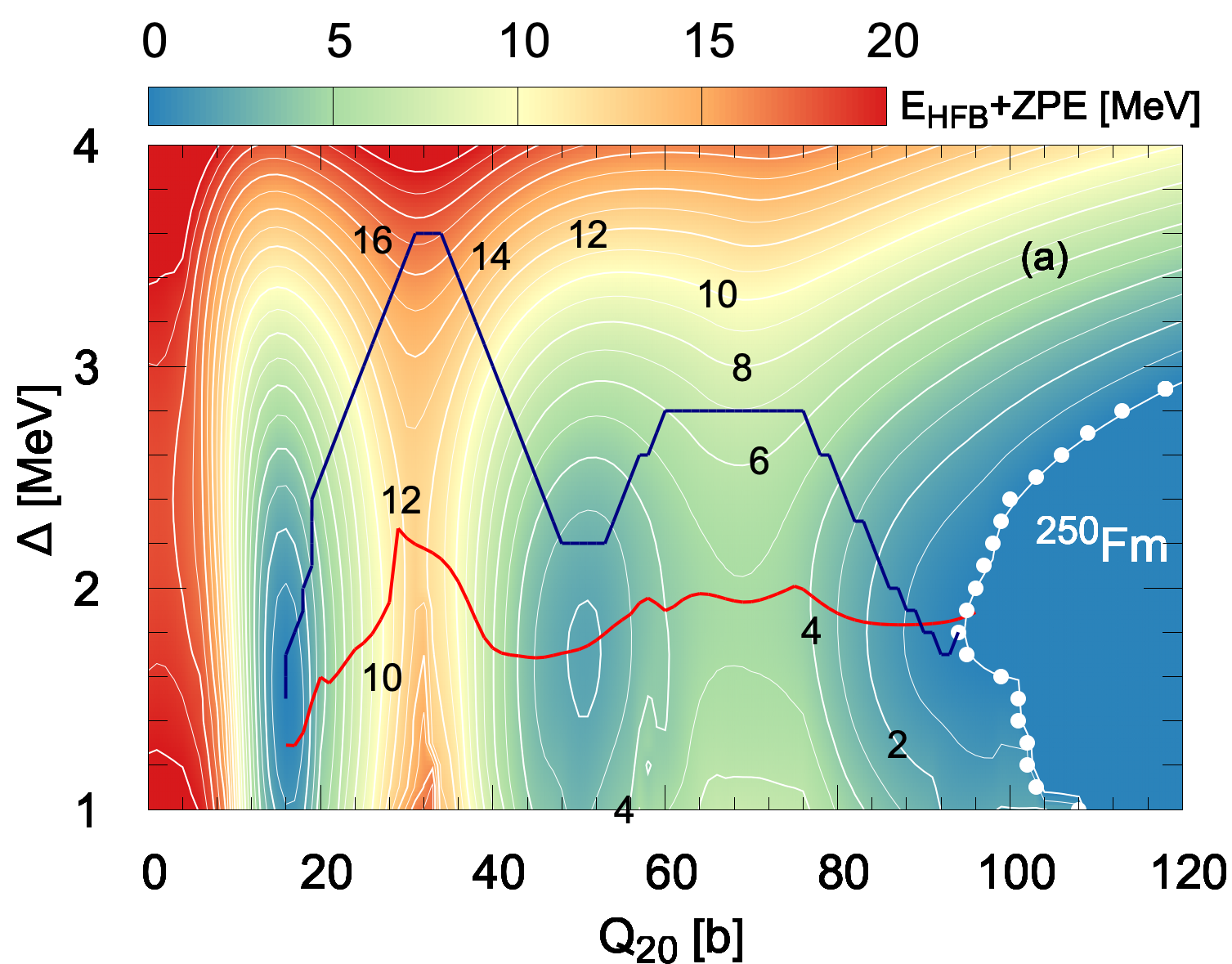}
\hskip-0.1cm
\includegraphics[scale=0.130,angle=0]{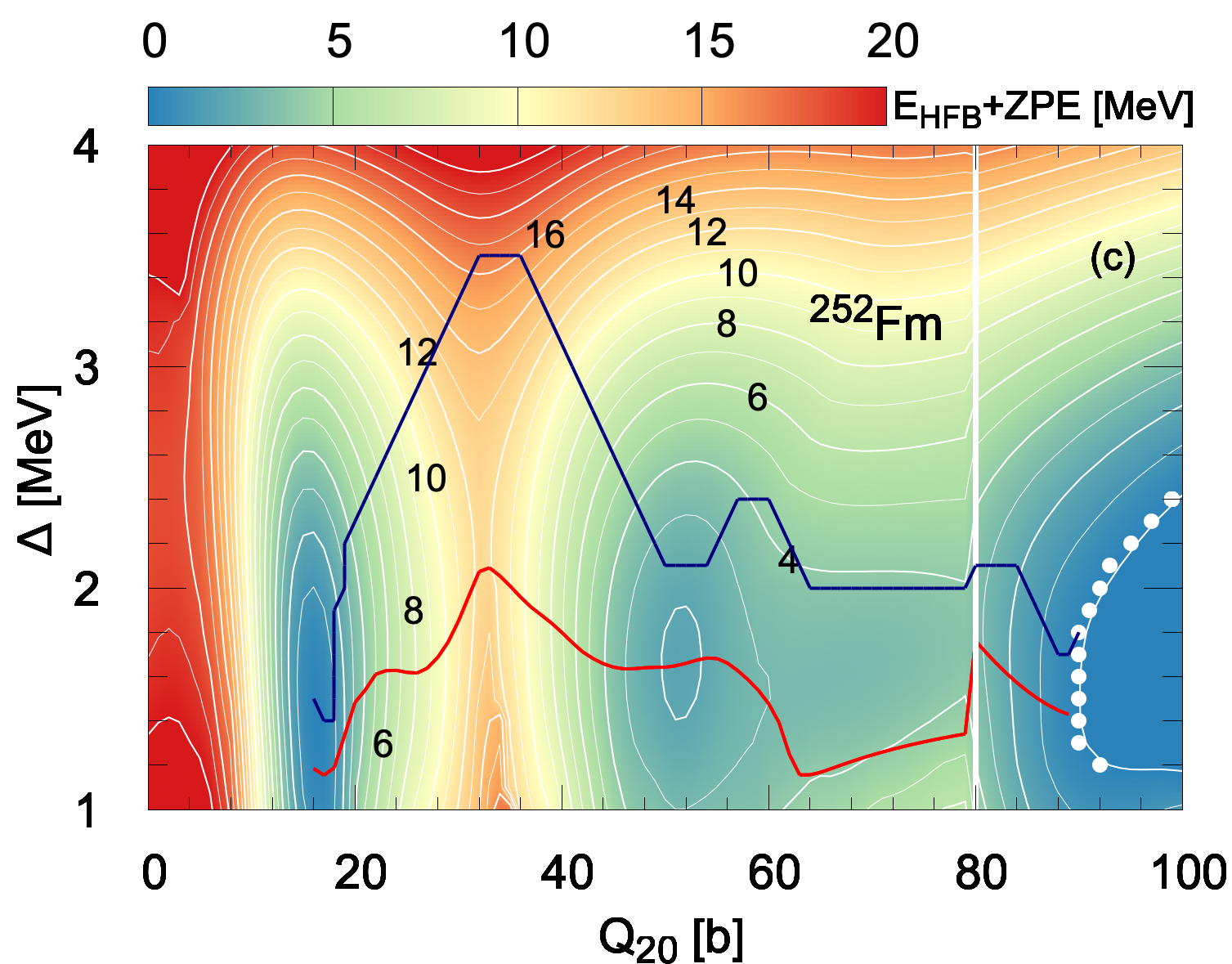}\\
\vskip-0.67cm
\includegraphics[scale=0.130,angle=0]{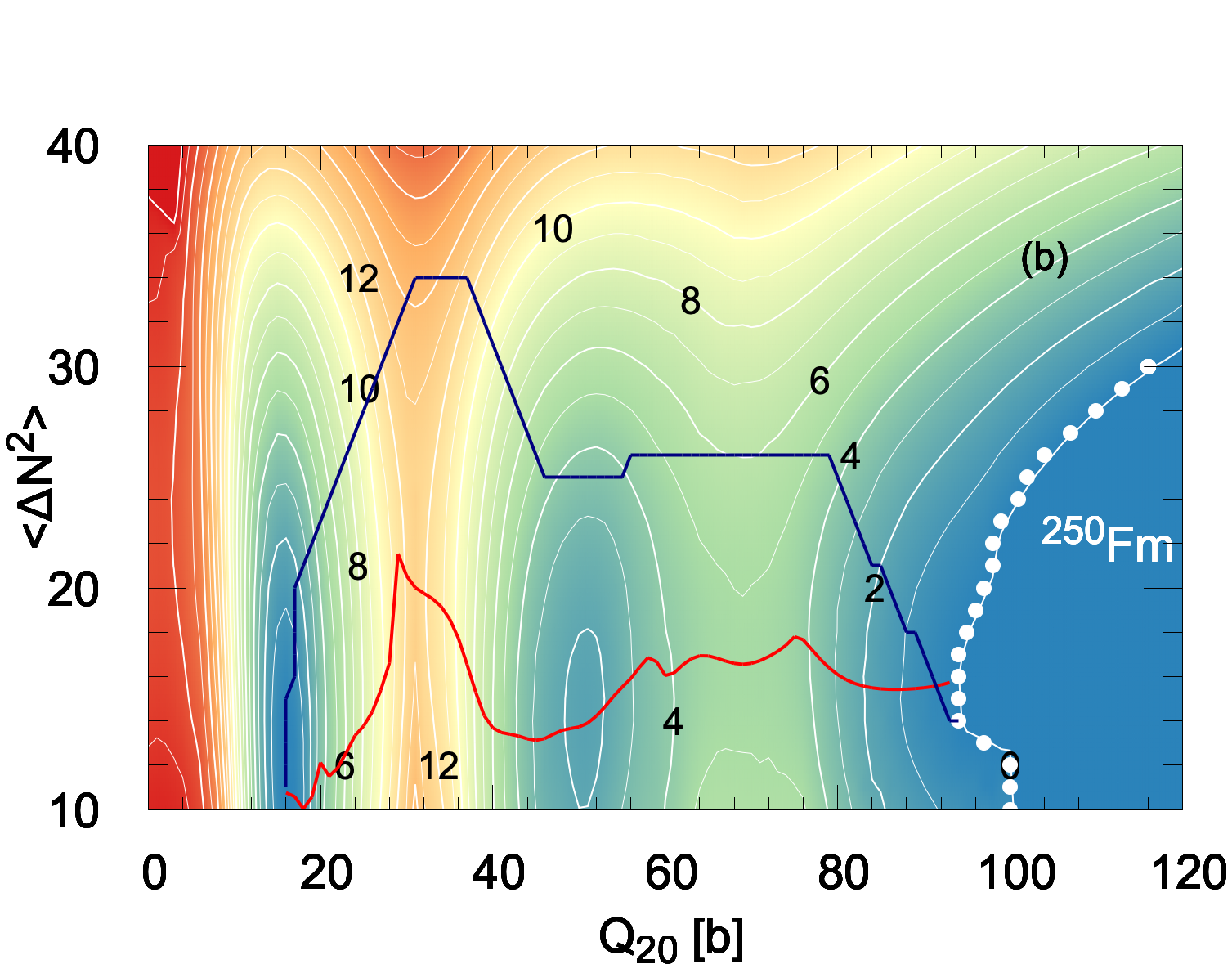}
\hskip-0.1cm
\includegraphics[scale=0.130,angle=0]{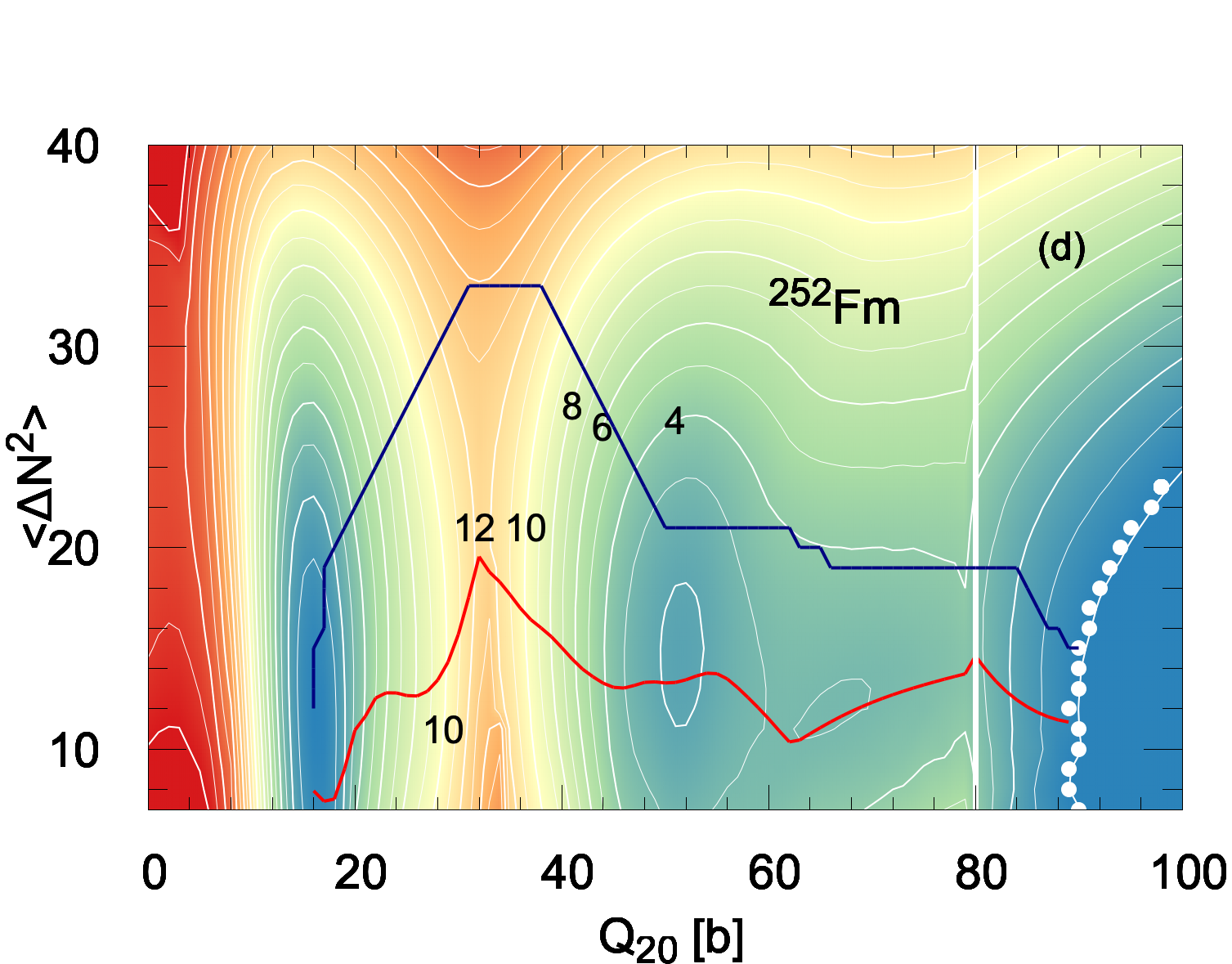}\\
\caption{Same as Fig.~\ref{dynFm1}, but for $^{250}$Fm and $^{252}$Fm.The white vertical solid line indicates the configuration change (see the text for the explanation).}
\label{dynFm3}
\end{figure*}
\begin{figure*}[!htb]
\includegraphics[scale=0.130,angle=0]{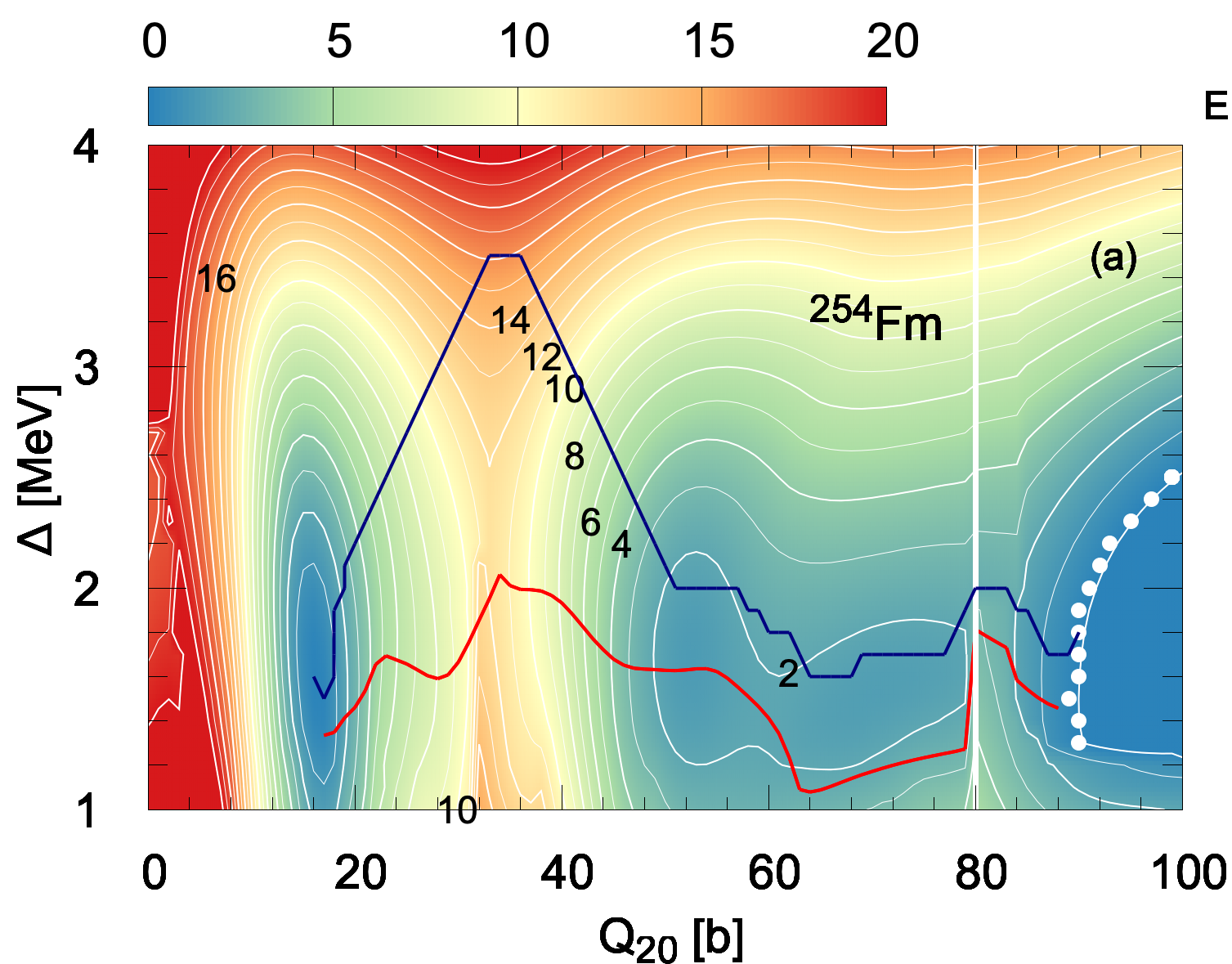}
\hskip-0.1cm
\includegraphics[scale=0.130,angle=0]{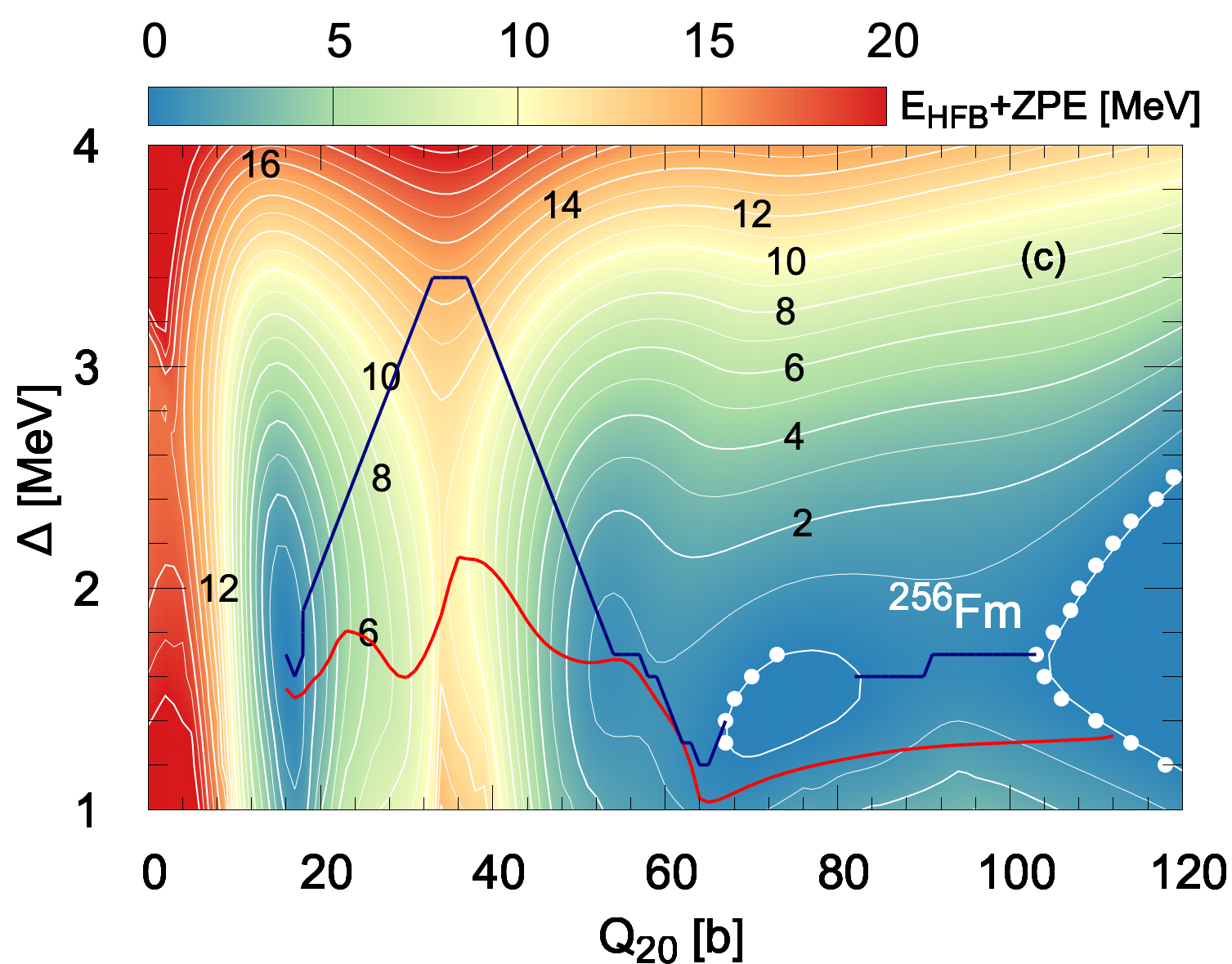}\\
\vskip-0.67cm
\includegraphics[scale=0.130,angle=0]{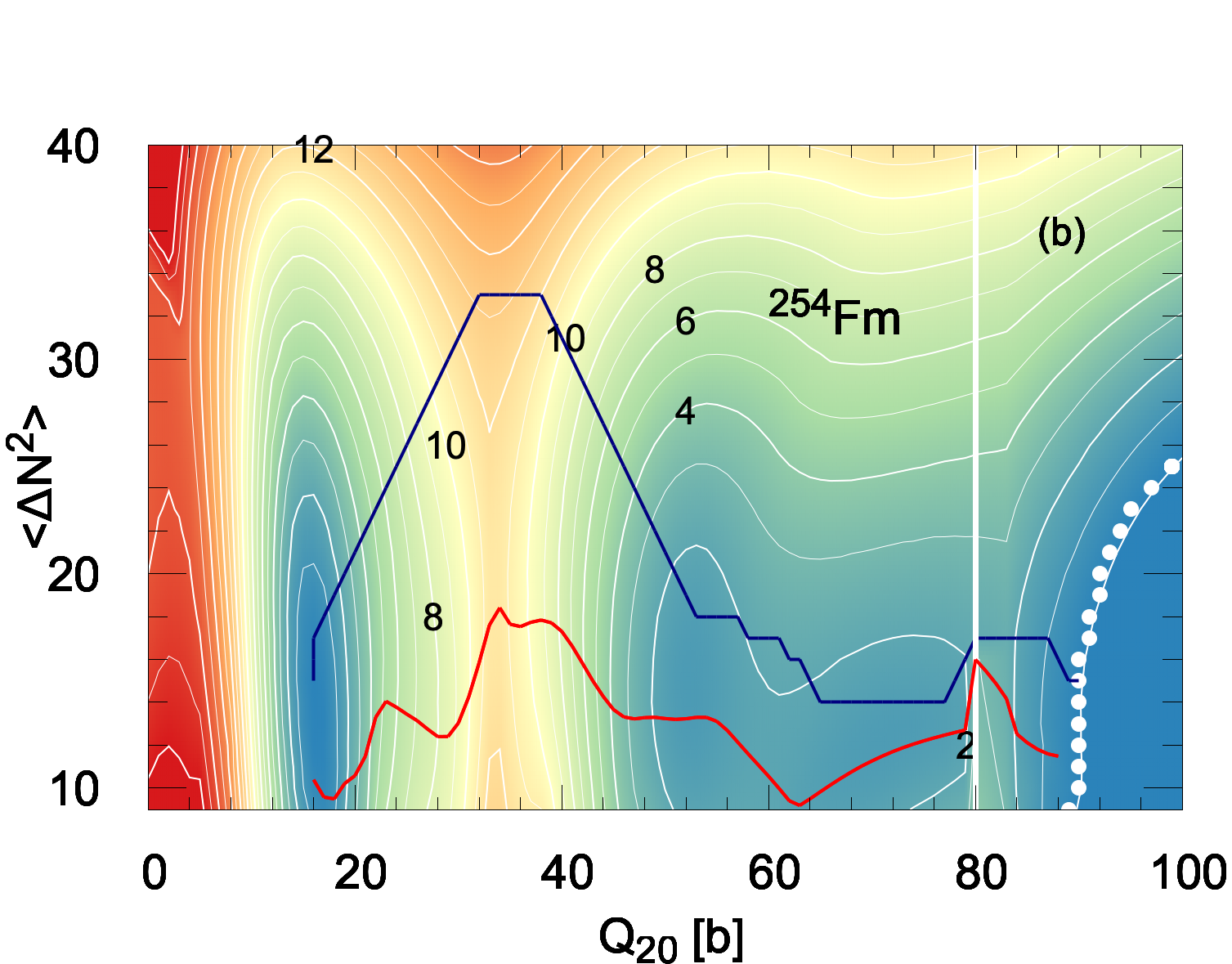}
\hskip-0.1cm
\includegraphics[scale=0.130,angle=0]{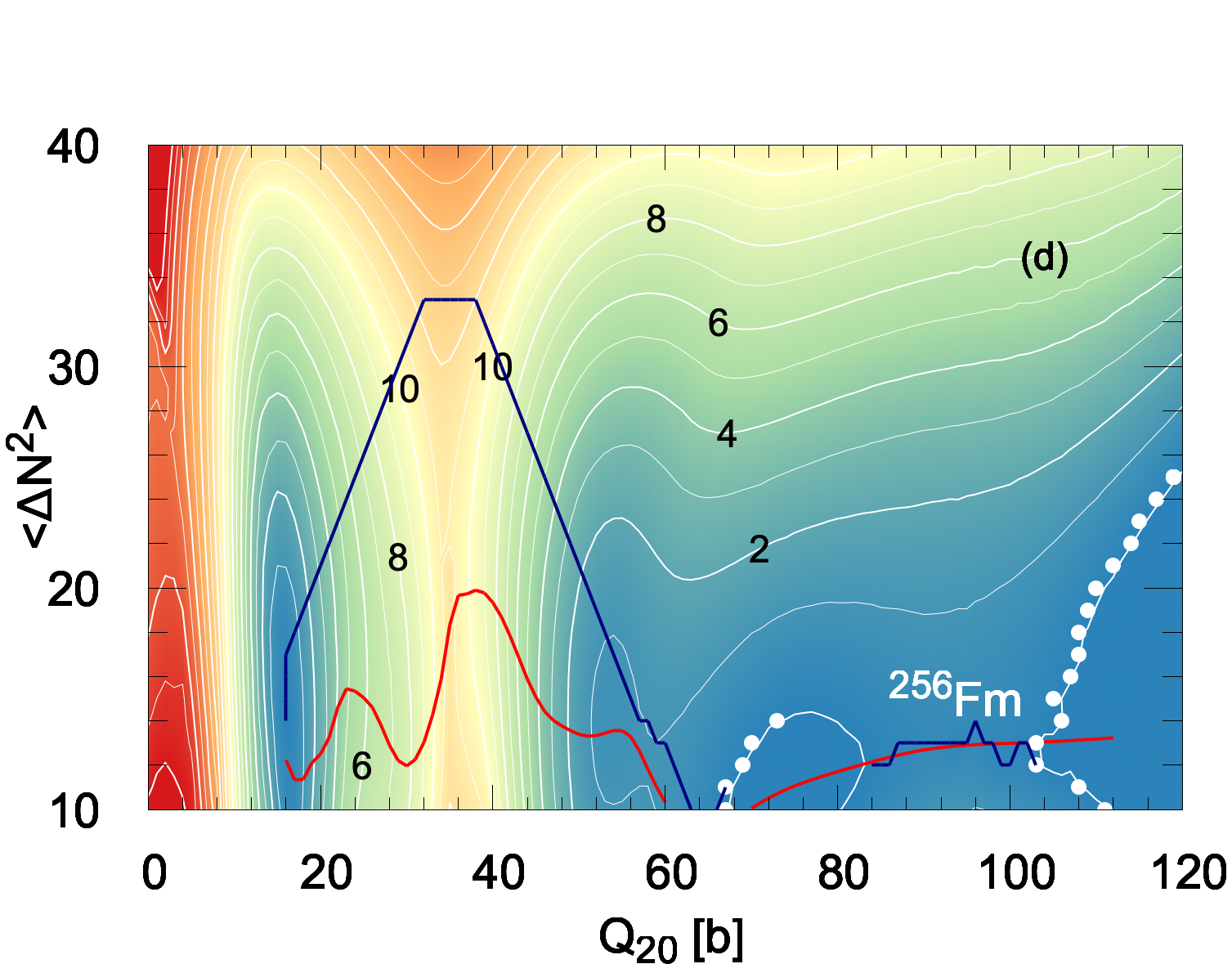}\\

\caption{Same as Fig.~\ref{dynFm1}, but for $^{254}$Fm and $^{256}$Fm.}
\label{dynFm4}
\end{figure*}
\begin{figure*}[!htb]
\includegraphics[scale=0.130,angle=0]{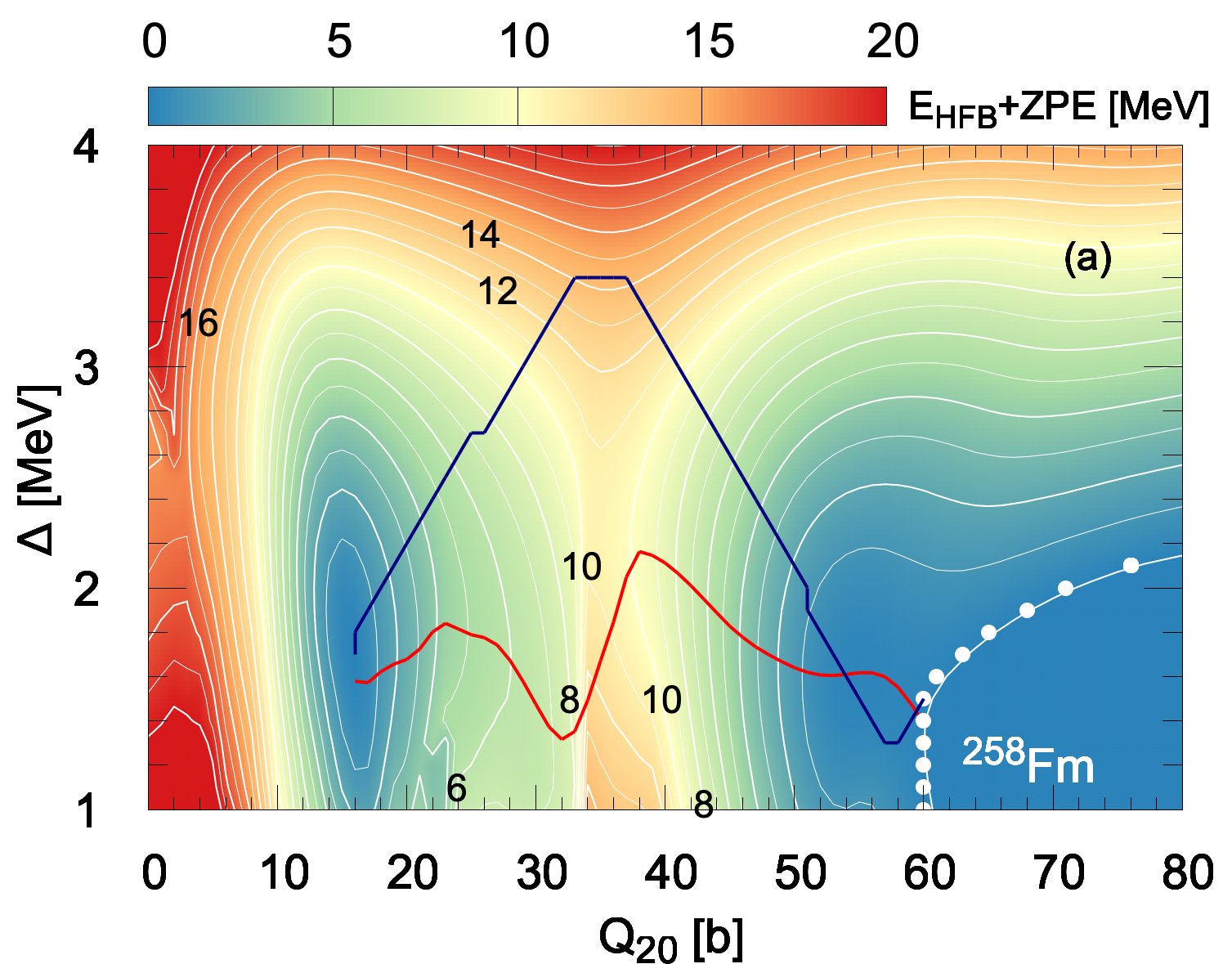}
\hskip-0.1cm
\includegraphics[scale=0.130,angle=0]{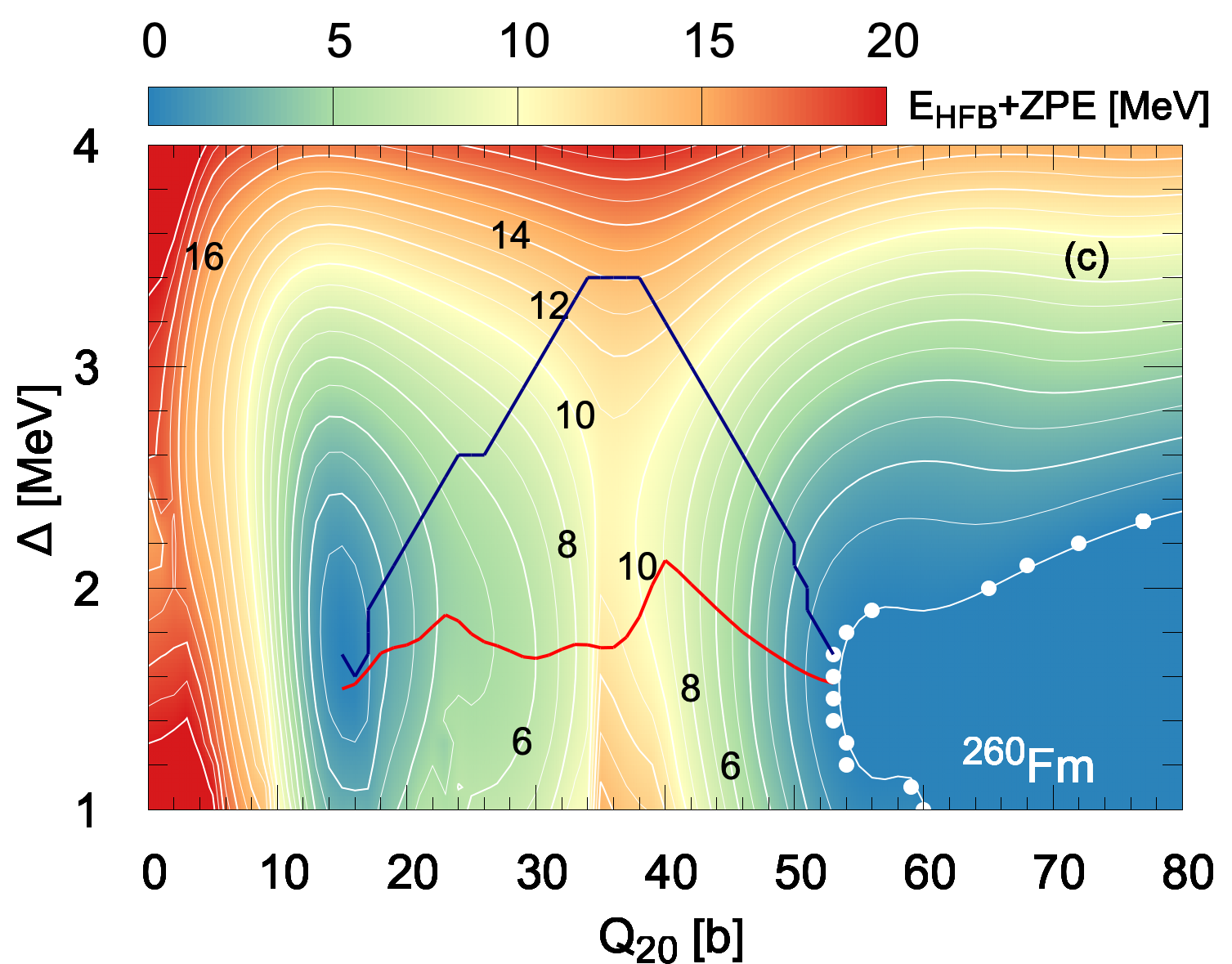}\\
\vskip-0.67cm
\includegraphics[scale=0.130,angle=0]{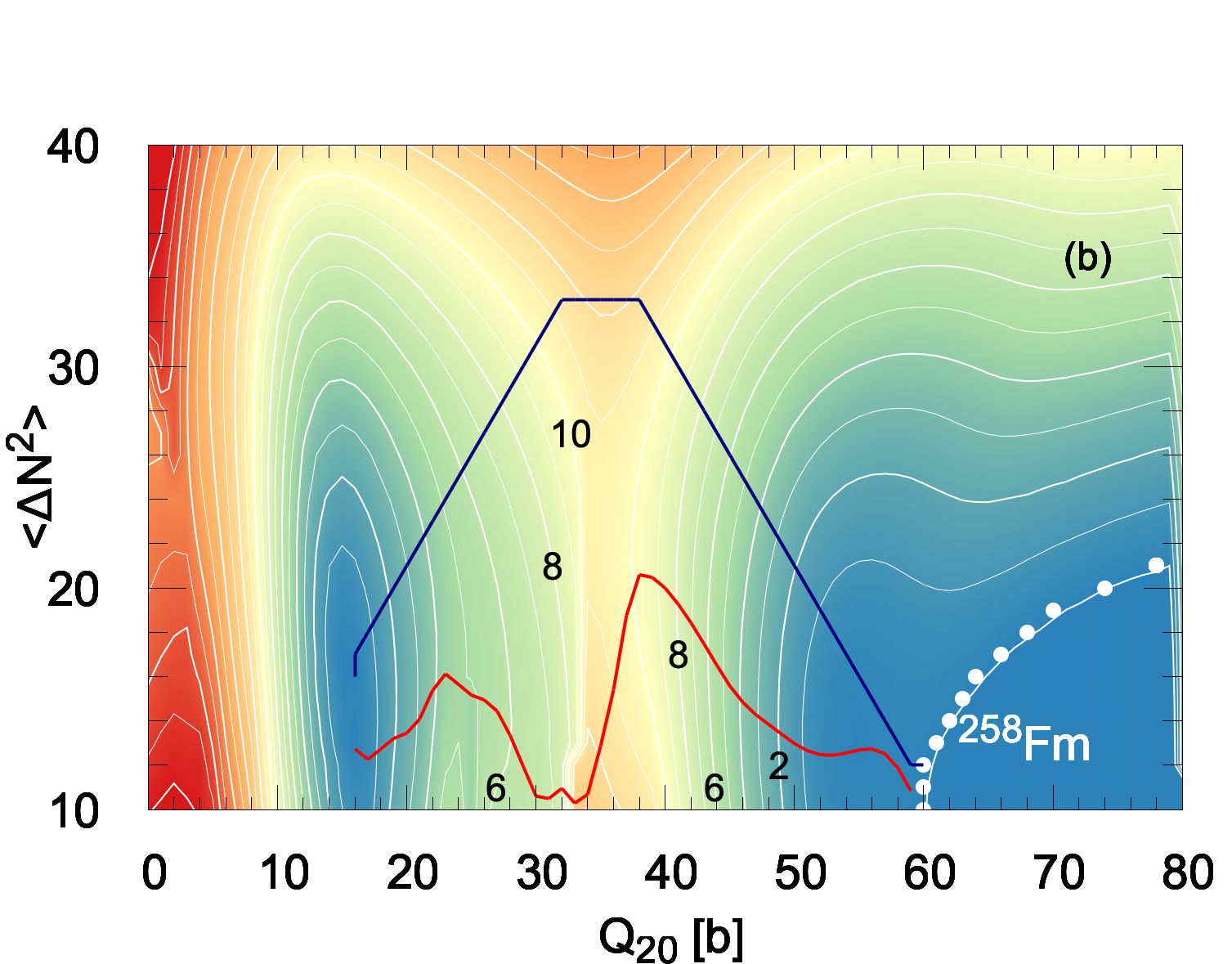}
\hskip-0.1cm
\includegraphics[scale=0.130,angle=0]{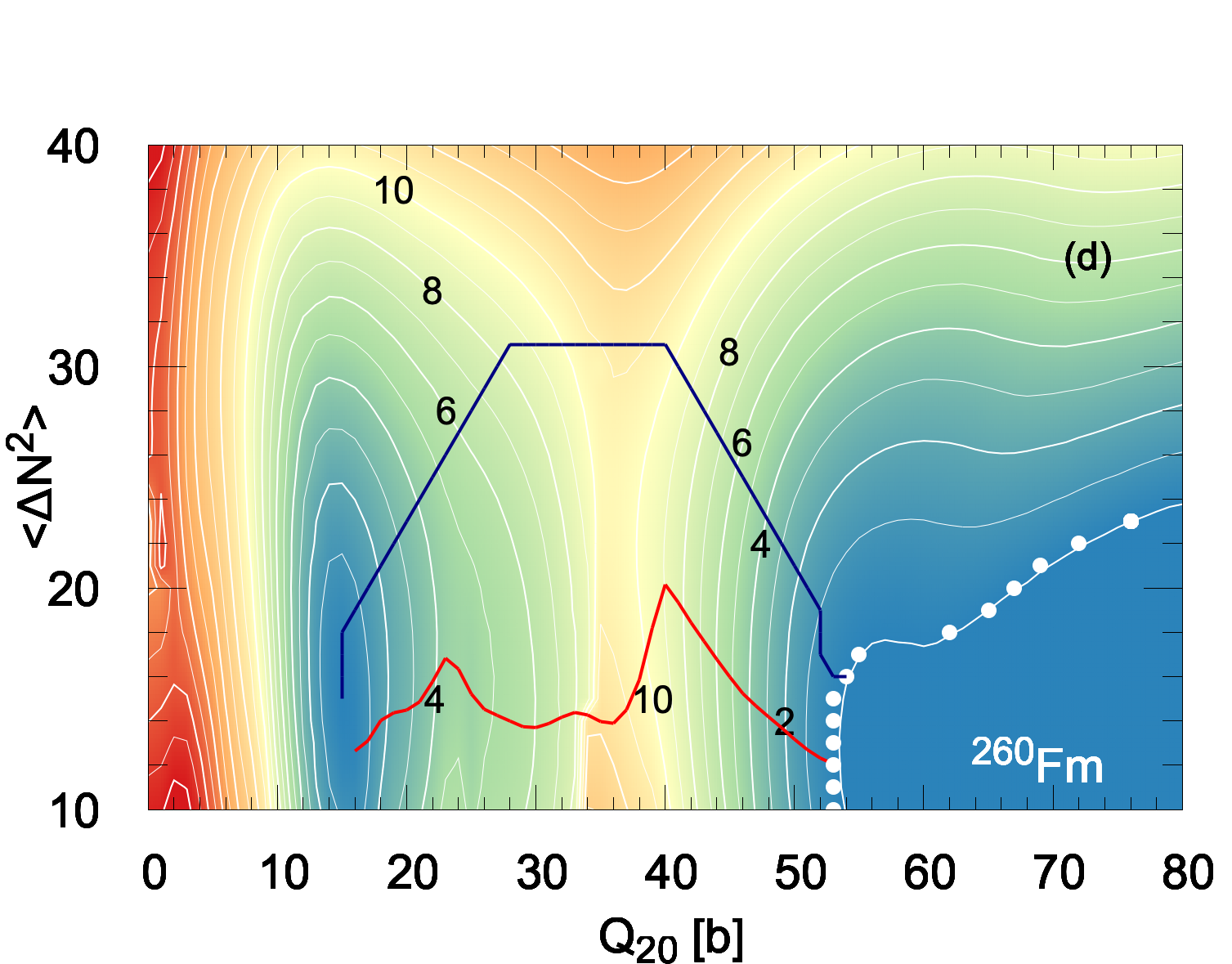}\\

\caption{Same as Fig.~\ref{dynFm1}, but for $^{260}$Fm and $^{262}$Fm. }
\label{dynFm5}
\end{figure*}
\begin{figure}[!htb]
\includegraphics[width=0.8\columnwidth, angle=0]{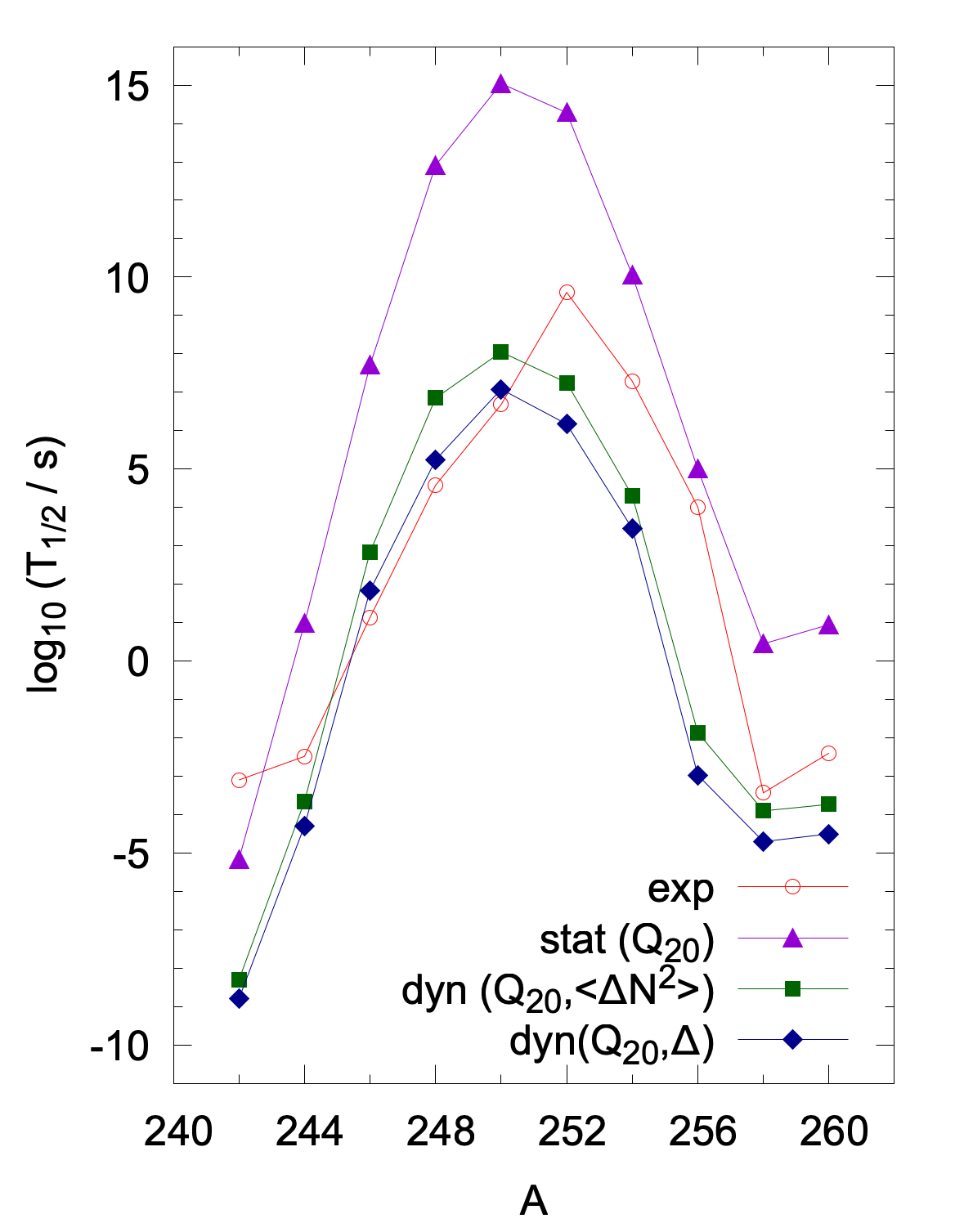}
\caption{Logarithms of the spontaneous fission half-lives of the fermium isotopes obtained within the dynamical (green 
squares and blue diamonds) and static (violet triangles) approaches compared with the experimental data (red circles) taken from~\cite{Wang_2021}.}
\label{fmt}
\end{figure}

\begin{figure*}[!htb]
\includegraphics[width=2\columnwidth, angle=0]{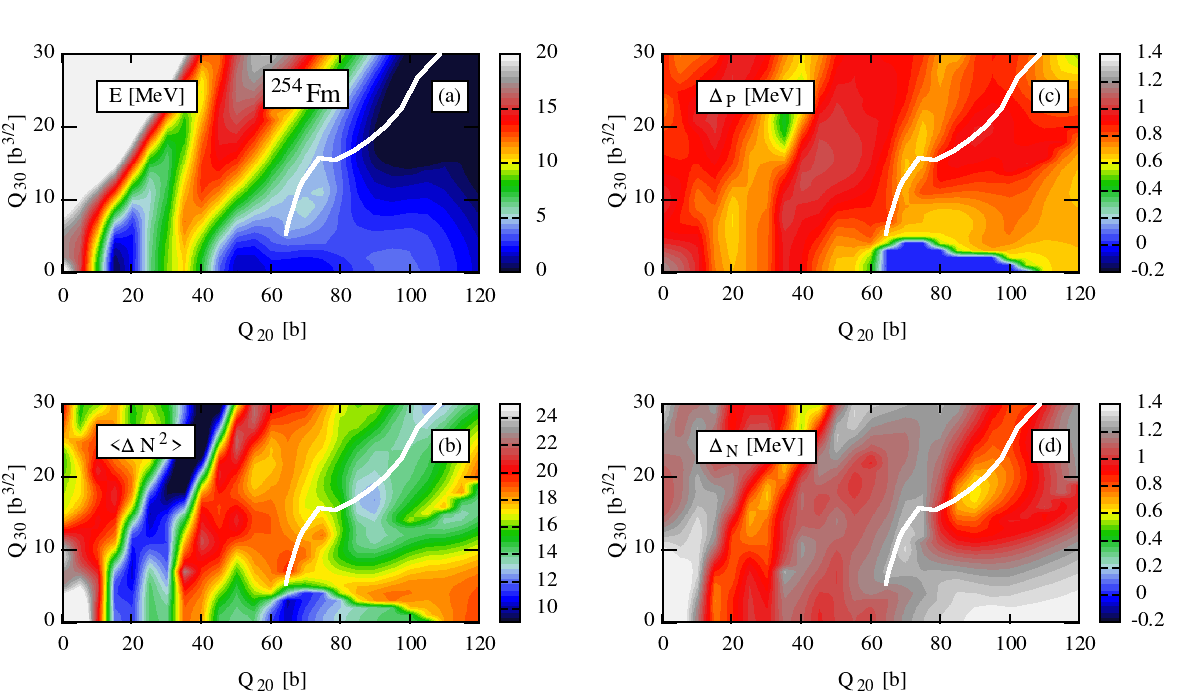}
\caption{(a) The PES of $^{254}$Fm in ($Q_{20}$, $Q_{30}$) plane. The white line shows the asymmetric fission path. (b) $\langle \Delta N^2 \rangle$, (c) proton $\Delta$ and (d) neutron $\Delta$ in the same space of deformations.}
\label{q2q3}
\end{figure*}

In this Section, we discuss the influence of pairing correlations on the dynamically determined fission paths. We use the pairing gap  $\Delta$ and average particle number fluctuations  
$\langle \Delta N^2 \rangle$ as alternative parametrizations of pairing correlations. The outcome of both approaches 
will be compared for $^{262}$Rf and the fermium isotopic chain

\subsection{Symmetric fission in $^{262}$Rf}
We start by studying the symmetric fission of $^{262}$Rf. This nucleus represents a convenient benchmark for our spontaneous fission studies due to the negligible role of octupole deformations, which ensures that the effects coming from the parity breaking do 
not obscure the overall picture of the role of pairing correlations. Also, this isotope has been extensively investigated in~\cite{zdeb2025} in the context of static fission paths. Here, we extend that previous analysis to the case of dynamical 
computations of spontaneous fission half-lives, including pairing as a collective degree of freedom. 

Figure~\ref{dynm} presents the PES of $^{262}$Rf in the ($Q_{20}$, $\Delta$)  and($Q_{20}$, $\langle 
\Delta N^2 \rangle$) planes, including the estimated least-energy 
and least-action paths. The PES and collective inertia presented in this section have been computed on a regular grid of 1~b for $Q_{20}$,  1 in $\langle \Delta N^2 \rangle$, 
and 0.1 MeV in $\Delta$. 
Evaluating the least-action integral (\ref{s}) involves all four inertia 
tensor components of the effective inertia (\ref{b_eff}) in the considered two-dimensional space. However, the quadrupole $B_{22}$ component usually dominates over the other terms
in the characterization of fission. 
It is shown in the right panels of Fig.~\ref{dynm}, respectively, in 
both spaces of collective variables. White isolines correspond to the potential energy in all four panels of 
Fig.~\ref{dynm}. We can see that in both coordinates, the dynamic fission path considerably diverges from the static 
one, especially in the region around the saddle point. For the $\langle \Delta N^2 \rangle$ ($\Delta$) constraint, the fission barrier height obtained with the dynamic fission path is around  2(5)~MeV higher than the saddle energy in the static case. Nevertheless, the action integral is much 
smaller as these paths avoid the region of high inertia around the top of the barrier.

The least-energy calculations performed in Ref.~\cite{zdeb2025} gives $\log_{10} (T_{1/2}/\mathrm{s})= 4.61$, which is much higher than the 
experimental value $\log_{10} (T_{1/2}/\mathrm{s})= -0.6$~\cite{Wang_2021}.
The spontaneous fission half-life along the least-action path is $\log_{10} (T_{1/2}/\mathrm{s})=-2.61$  when $\Delta$ 
is involved as a dynamic coordinate, and $\log_{10} (T_{1/2}/\mathrm{s})=-1.27$ with $\langle \Delta N^2 \rangle$. These 
results show a considerable improvement in comparison with the least-energy calculations. The gain in tunneling probability obtained in the dynamical calculations can be easily understood from Fig.~\ref{1d}. The least-action paths increase the energy along the barrier by no more than 50\%.
At the same time, the $B_{22}$ component of the inertia is reduced by a factor of 20. Consequently, the product of energy and effective inertia is lower all along the fission path for dynamical calculations, and the action entering Eq.~(\ref{s}) is reduced. 
A variation of 1.3 orders of magnitude between the half-life predicted by dynamical calculations can be noticed. This
difference is related to the discretization of the PES and collective inertia into a regular grid and falls within the typical uncertainty of theoretical estimations of spontaneous fission lifetimes. One can see that the path found on the ($Q_{20}$,$\Delta$) PES reaches around 3 MeV higher energy, but the impact on the action integral is compensated by much lower inertia in the same region  in the ($Q_{20}$, $\langle \Delta N^2 \rangle$) space.
Slight deviations in the dynamical evaluation of the path, e.g., in the region of the second minimum around $Q_{20}= 55$ b, are responsible for higher action in the calculations in ($Q_{20}$, $\langle \Delta N^2 \rangle$) space.

\subsection{Fermium isotopic chain}

The spontaneous fission of the fermium isotopic chain is particularly interesting due to several unique  
decay properties. Fermium isotopes exhibit a sharp transition from spontaneous fission as the dominant 
decay mode to alpha decay. Lighter isotopes, $^{244 - 248}$Fm, undergo almost pure spontaneous fission, while 
heavier ones like $^{250 - 260}$Fm show a competition between SF and alpha decay~\cite{Kondev_2021}. Spontaneous fission half-lives in the fermium isotopic chain create a characteristic inverted parabola connecting   
short-lived $^{242}$Fm  ($T_{1/2}=800 \;\mu$s) and $^{258}$Fm ($T_{1/2}=370 \;\mu$s) with the long-lived $^{252}$Fm ($T_{1/2}=129$~y). Another point of interest in this isotopic chain is the transition from asymmetric to 
symmetric fission. Most of the lighter fermium isotopes ($^{244-256}$Fm) undergo asymmetric fission, with mass distributions 
favoring heavy fragments around the magic numbers $Z=50$ and $N=82$  (tin-like isotopes). In $^{258}$Fm 
bimodal fission  has been observed~\cite{Hulet1986}. It is a transition nucleus towards the symmetric mode in the heaviest fermium isotopes. 
Fission in the chain of fermium isotopes was studied in many theoretical papers \cite{NURMIA196778,BARAN1978,warda2002,baran2,Bonneau2004,
Hooshyar2005,Warda_2006,baran3,PhysRevC.79.064304,PhysRevC.87.024320,Giuliani2013,baranstaszczak,PhysRevC.97.034319,Baran2015,PhysRevC.98.034308,PhysRevC.105.034604}. Still, the fermium isotopic chain remains a crucial 
case study and challenging for any modern nuclear theory.

We have investigated the chain of even-even fermium isotopes from $A=242$ to 260. The two-dimensional PESs with $\langle 
\Delta N^2 \rangle$ and $\Delta$ as a pairing degree of freedom are 
presented in Figs.~\ref{dynFm1} - \ref{dynFm5}. Several characteristic 
types of the PES can be found in this series, allowing for the 
examination of the influence of pairing correlation in various types of 
fission. The first group comprises the two lightest isotopes 
$^{242-244}$Fm. Their PESs look very similar in both coordinates.
They are characterized by a well-defined inner fission barrier of 8 MeV 
height, followed by a 1-2~MeV narrow outer fission barrier. In 
$^{242}$Fm, a second fission isomer with a short second outer barrier 
is also present, which disappears in $^{244}$Fm. 
To compute the half-lives in these cases where two barriers are present
one should take into account the possibility of enhanced transmission
coefficients due to the presence of resonances associated to virtual
states in the fission isomer well \cite{RevModPhys.52.725}. Those resonances
are usually very narrow as a function of the energy and therefore their
impact in the final results is very limited unless the excitation energy
coincides with the energy of the resonance. Therefore, we have decided
not to take into account its effect in the present calculation. In Ref \cite{IGNATYUK1969} 
the problem of the transmission probability in the presence of two wells
was addressed in the framework of the WKB approximation (and constant inertias) and a very 
elaborated formula depending on the energy dependent action was proposed.
After an average on excitation energy the formula reduces to $T=T_{A}T_{B}/(T_{A}+T_{B})$
which coincides with the expression for a potential well made of parabolas and a constant inertia \cite{Oberstedt2023}.
The formula in \cite{IGNATYUK1969} has been criticized by several authors 
(see, for instance \cite{FROMAN1970,PhysRevC.47.2801}) and, according to Ref \cite{Oberstedt2023}
is only meaningful when the excitation energy is larger than the height of the
lowest barrier, which is not the case in the present calculation. In our
case, as $T_{B} >> T_{A}$, the formula of \cite{IGNATYUK1969}  gives
$T=T_{A}$ which completely neglects the impact of the second barrier in the transmission
probability. For the above reasons, we prefer to compute the transmission
coefficient as the simple product of the two transmission coefficients.

In the $^{242-244}$Fm isotopes, the calculated half-lives are very low, even 
shorter than experimental data, see Fig.~\ref{fmt}. The least-action 
paths have the same trend as in $^{262}$Rf and reach higher energies 
with stronger pairing correlations. Slightly higher energies are 
achieved at the barrier for the $\Delta$ coordinate. Shorter half-lives 
along the least-action path give a worse agreement with experimental 
data than the least-energy path for $^{242}$Fm.

In  $^{246-250}$Fm isotopes, the topography of the 
PES considerably changes. Both the inner and outer fission barriers become wider and higher, reaching a 10 MeV and 7 MeV height, respectively.
The outer barrier corresponds to the reflection-asymmetric shapes of a 
nucleus. It is important to mention that the fission path with octupole deformation plays a crucial role in reproducing the asymmetric fission fragment mass distribution in the central part of the fermium isotopic chain. The visible discontinuity of the PES on the second fission barrier is related to the switching on of octuple deformation, 
as discussed in Ref.~\cite{zdeb2021}. A more detailed study would require analysis in 3-dimensional space, 
including a constraint on the octupole moment. The fission barrier of the mid-mass Fm isotopes extends to large deformations $Q_{20}=80 - 100$~b. As a consequence, the fission half-lives of these nuclei are much longer.
In these nuclei, the least-energy fission path overestimates measured half-lives by around five orders of magnitude, while experimental data are almost perfectly reproduced with the least-action fission path.

In heavy Fm isotopes, starting from $^{252}$Fm, an unexpected deviation from the smooth behavior of the pairing interaction can be noticed in the symmetric fission valley at large deformations. This region of the PES corresponds to the symmetric compact fission mode~\cite{warda2002}, when a nucleus splits into two identical, spherical, semi-magic tin isotopes. Already in the second minimum, a nucleus takes the shape of two partially overlapping spheres. In this fission valley, between $Q_{20}= 64-95$ b, the proton pairing gap $\Delta_p$ drops to zero, as shown in Fig.~\ref{q2q3} for the representative example of $^{254}$Fm. At the same time, neutron $\Delta$ and $\langle \Delta N^2 \rangle$ are unaffected by any substantial modifications. The extraordinary behavior of $\Delta_p$ can be explained by the huge gap, between 2 and 3 MeV, in the spectrum of the proton single-particle levels at the Fermi level. This resembles the structure of a magic nucleus. Such a huge gap causes the occupancies of the proton orbitals to be no longer smooth, taking one or zero values. In consequence, the calculated $\Delta_p$ parameter diminishes. 

The collapse of the proton pairing affects the linear relationship between $\Delta=\Delta_p+\Delta_n$ and $\langle \Delta N^2 \rangle$.
This is because when $\Delta_p$ vanishes, the total gap parameter is equal to its neutron component only. Due to the missing proton component, the total gap parameter in the region is roughly 20\% lower than the value predicted by the linear fit to $\langle \Delta N^2 \rangle$. In other words, the linear relationship between $\Delta$ and $\langle \Delta N^2 \rangle$ breaks down in the regime of pairing collapse, such as the compact symmetric fission path of $^{252}$Fm.

The static fission path, corresponding to the largest tunneling probability, assumes a relatively large jump from reflection symmetric to octupole deformed shapes, and the PES in these two isotopes is not continuous. The right-hand part of the PES from the white vertical line describes fission with pronounced mass asymmetry. In these two nuclei, dynamical treatment of pairing significantly reduces fission half-lives to the level below measured values.

The fission pathway of the isotope $^{256}$Fm proceeds through a shape isomer and ends around $Q_{20}$ = 100 b. On the PES of the heaviest isotopes $^{258-260}$Fm, one can see that the second minimum is below the ground state, and the second turning point of the barrier is again around $Q_{20}=60$ b, as in the case of the lightest isotopes. Thus, the region of collapse of proton pairing is beyond the barrier. The half-lives calculated with the pairing degrees of freedom are much shorter than the static ones. In the case of $^{258-260}$Fm, these results are much closer to the experimental observables. 

For all the studied Fm isotopes presented in Figs. \ref{dynFm1} - \ref{dynFm5} we can see that the least-action paths lead through 
the regions of the PES with stronger pairing interaction than in the ground state or along the least-energy path. 
This trend is visible both in the first, symmetric and the second, asymmetric hump of the barrier.
Despite a higher energy barrier, the reduction of the effective collective inertia causes a reduction of the action integral 
and an increase in the tunneling probability.
The least-action path reaches slightly higher energies in the  $\Delta$ PES than in the one with $\langle \Delta N^2 
\rangle$ as a collective coordinate. It is also worth mentioning that in most cases, the dynamic exit point deviates 
from the static one in the presented hyper-space. Nevertheless, this difference does not sensibly affect the shape of a nucleus 
and its reflection asymmetry.

The spontaneous fission half-lives along the fermium isotopic chain depicted in Fig.~\ref{fmt} show a significant impact of the pairing degree of freedom on the theoretical predictions. The half-lives are reduced by three to eight orders of magnitude compared to the calculations along the least-energy paths. The inverted parabola is shifted down. In many isotopes ($^{246-252}$Fm, $^{258-260}$Fm) this shift considerably improves reproduction of the experimental data. However, for $^{242}$Fm nucleus where the static fission path underestimates the spontaneous fission lifetimes, the dynamic calculation does not improve the agreement with experimental data. This result may indicate the lack of relevant collective degrees of freedom related to higher multipole moments, or the limitation of the perturbative calculation of collective inertias~\cite{PhysRevC.84.054321,Giuliani2018b}. Apart from those discrepancies, we observe a nice reproduction of the trend with neutron number. Some quantitative discrepancies still remain that can be associated with the assumptions made in the calculation: validity of the WKB approximation, the approximations used in the evaluation of the inertias, the determination of the $E_0$ parameter, the impact of the different constituents of the effective interaction, the collective degrees of freedom considered, etc. Most of the just-mentioned effects are discussed to some extent in the white paper \cite{Bender_2020} and represent an exciting field of research for the future.

\section{CONCLUSIONS}
We studied the impact of the pairing degree of freedom on the dynamical description of the fission process. We have applied the HFB theory with the Gogny D1S force to describe the PES and collective inertia of several fissioning nuclei. The calculations have been performed in a hybrid two-dimensional deformation space with constraints on the quadrupole moment and pairing correlations. The latter has been constrained employing either the pairing gap parameter $\Delta$ or the average particle number fluctuations $\langle \Delta N^2 \rangle$.

The Dijkstra algorithm has been applied to determine the least-action fission path on the two-dimensional PES with the effective inertia. The dynamical fission paths have been traced for $^{262}$Rf and the chain of fermium isotopes. Both collective degrees of freedom associated with pairing lead to comparable results. We find that in all the calculations, the highest tunneling probability occurs when the path passes through regions of higher energy than the static saddle point, due to a significant reduction in collective inertia. This is consistent with the pioneering works in this field.

The one-body approximation of the collective inertias associated with the $\langle \Delta N^2 \rangle$ constraint works well along fission paths of the studied isotopes. Calculations performed in a space with  $\langle \Delta N^2 \rangle$ as a pairing degree of freedom give slightly longer half-lives, around one order of magnitude larger than those with $\Delta$ in all considered isotopes. This systematic shift originates from the assumed linear relationship between $\Delta$ and $\langle \Delta N^2 \rangle$. The finite resolution of the mesh grid and small deviations from linearity in the range $10\leq \langle \Delta N^2 \rangle \leq 40$ spoil the one-to-one correspondence between the PESs obtained with both constraints. In consequence, the energies in the $(Q_{20},\Delta)$ PES are a slightly overestimated compared to the $(Q_{20},\langle N^2 \rangle)$ results. Thus, the dynamical paths give higher energy and, at the same time, smaller collective inertia along the fission path, resulting in shorter half-lives. A more accurate fit, obtained by taking into account only the relevant range $15\leq \langle \Delta N^2 \rangle \leq 30$ and/or separate fits for protons and neutron components, should reduce these discrepancies. 

The least-action fission paths have been determined along the fermium isotopic chain. Considerable improvement in reproducing the spontaneous fission half-lives data is achieved in most cases compared to the static approach. It applies both to symmetric and asymmetric fission modes. This confirms the importance of including the pairing degree of freedom in the description of fission dynamics.

The fermium isotopic chain provides a rare, experimentally accessible case of a spontaneous fission transition from 
asymmetric to symmetric modes.  The transition highlights the need to understand how the 
fission landscape evolves when dynamical pairing is considered along with the octupole moment to drive the asymmetric to symmetric transition. The presented results constitute a promising direction for future research. This approach, coupled with advanced nuclear
models and numerical techniques, promises to provide new insights into the role of pairing in the fission dynamics of
heavy nuclei, ultimately refining our understanding of the fission process and aiding in the prediction of fission
half-lives in super-heavy nuclei, as well as an improvement of the fission fragment mass yields studies.

\section{ACKNOWLEDGEMENTS}
The work of A.Z. and M.W. is part of the project No. 2021/43/P/ST2/03036, co-funded 
by the National Science Centre and the European Union’s Horizon 2020 
research and innovation programme under the Marie Skłodowska-Curie grant 
agreement no. 945339. The work of L.M.R. and S.A.G. is supported by Spanish Agencia Estatal de Investigacion (AEI) of the Ministry of Science and Innovation (MCIN) under Grant No. PID2021-127890NB-I00. S.A.G. acknowledges support funded by MCIN/AEI/10.13039/501100011033 and the ``European Union NextGenerationEU/PRTR'' under Grant Agreement No. RYC2021-031880-I.

\section{DATA AVAILABILITY}
The data that support the findings of this article are openly available \cite{oa}.

\bibliographystyle{apsrev4-1}
\bibliography{references}


\end{document}